\documentclass[aps,prx,reprint,superscriptaddress]{revtex4-1}
\usepackage{amsfonts}
\usepackage{amsmath}
\usepackage{amssymb}
\usepackage{graphicx}
\usepackage{epstopdf}
\usepackage{color}
\usepackage{bbold}
\usepackage{bm}
\usepackage[dvipsnames]{xcolor}
\usepackage{pifont} 

\renewcommand{\vec}[1]{\boldsymbol{#1}}

\begin{document}

\title{Theory of Spin Hall Magnetoresistance from a Microscopic Perspective}


\author{X. P. Zhang}
\email{zianpeng_zhang001@ehu.eus}
\affiliation{Donostia International Physics Center (DIPC), Manuel de
Lardizabal, 4. 20018, San Sebastian, Spain}
\affiliation{Centro de Fisica de Materiales (CFM-MPC), Centro Mixto CSIC-UPV/EHU,
20018 Donostia-San Sebastian, Basque Country, Spain}

\author{F. S. Bergeret}
\email{fs.bergeret@csic.es}
\affiliation{Donostia International Physics Center (DIPC), Manuel de
Lardizabal, 4. 20018, San Sebastian, Spain}
\affiliation{Centro de Fisica de Materiales (CFM-MPC), Centro Mixto CSIC-UPV/EHU,
20018 Donostia-San Sebastian, Basque Country, Spain}

\author{V. N. Golovach}
\email{vitaly.golovach@ehu.eus}
\affiliation{Donostia International Physics Center (DIPC), Manuel de
Lardizabal, 4. 20018, San Sebastian, Spain}
\affiliation{Centro de Fisica de Materiales (CFM-MPC), Centro Mixto CSIC-UPV/EHU,
20018 Donostia-San Sebastian, Basque Country, Spain}
\affiliation{IKERBASQUE, Basque
Foundation for Science, 48013 Bilbao, Basque Country, Spain}

\begin{abstract}
We present a theory of the spin Hall magnetoresistance of metals in contact with magnetic insulators.
We express the spin mixing conductances, which govern the phenomenology of the effect, in terms of the microscopic parameters of the interface 
and the spin-spin correlation functions of the local moments on the surface of the magnetic insulator.
The magnetic-field and temperature dependence of the spin mixing conductances leads to a rich behaviour of the resistance due to an interplay between the Hanle effect and spin mixing at the interface.
Our theory provides a useful tool for understanding the experiments on heavy metals in contact with magnetic insulators of different kinds, and it predicts striking behaviours of the magnetoresistance. 
\end{abstract}

\maketitle

\emph{Introduction}-
The spin-orbit coupling (SOC) in metals and semiconductors leads to a conversion between the charge and spin currents, 
which results in the spin Hall effect (SHE) and its inverse effect~\cite{DyakonovPerel1971JETPL,DyakonovPerel1971PLA,hirsch1999spin,sinova2004universal,valenzuela2006direct,kimura2007room,kato2004observation,sih2005spatial,wunderlich2005experimental,zhou2018observation,maekawa2017spin,sinova2015spin,nakayama2016rashba}. 
A manifestation of the SHE in a normal metal (NM) is a modulation of the magnetoresistance (MR) with respect to the direction of the applied magnetic field when the metal is in contact with a magnetic insulator (MI) in  NM/MI structures~\cite{isasa2016spin,althammer2013quantitative,huang2012transport}. This effect, called the spin Hall magnetoresistance (SMR),  has been observed in several experiments~\cite{weiler2012local,nakayama2013spin,avci2015unidirectional,hahn2013comparative,dejene2015control}.  The origin of the  SMR is the  spin-dependent scattering at the NM/MI interface  which depends on the angle between the polarization of spin Hall current  and the  magnetization of the MI~\cite{nakayama2013spin,chen2013theory}. The latter can be controlled by magnetic fields.

Although the theory of SMR~\cite{nakayama2013spin,chen2013theory}  is well established and provides a qualitative description of the effect, it does not describe the dependence of the resistivity on the strength of the applied magnetic field $B$, nor on the temperature $T$.
The spin mixing conductances, which are at the heart of the SMR effect,
have traditionally been regarded as phenomenological parameters in every experiment, because their computation was thought to be a formidable task which could only be carried out by \emph{ab initio} methods~\cite{jia2011spin,carva2007spin,zhang2011first,xia2002spin,dolui2019spin}.
Recent experiments~\cite{meyer2014temperature,velez2018spin,das2019temperature}
show, however, that the SMR effect depends both on $B$~\cite{velez2018spin} and on $T$~\cite{meyer2014temperature,velez2018spin,das2019temperature},
and that the magnetic state of the MI plays an important role for SMR.
Furthermore, the magnetic field alone leads  
to the Hanle magnetoresistance (HMR)~\cite{dyakonov2007magnetoresistance,velez2016hanle}, 
which has an identical angular dependence to SMR~\cite{velez2016hanle}, but is not requiring an MI.
Despite the fact that SMR and HMR have different origins, they cannot always be easily separated in experiments, 
which adds onto the uncertainties of interpreting the experimental data.
It is, therefore, desirable to have a theory of SMR 
which has predictive power about the dependence of the spin mixing conductances on $B$ and $T$ and is able to cover a wide range of magnetic
system, from classical to quantum magnets.

In this  letter, we present a general  theory of the electronic  transport in NM/MI structures. 
We describe the spin-dependent  scattering at the NM/MI interface via a microscopic model based on the \textit{sd} coupling between local moments on the MI surface and itinerant electrons in the NM. The temperature and magnetic-field dependence of the interfacial scattering coefficients is obtained by expressing them in terms of spin-spin correlations functions. The latter are determined by the magnetic behavior of the MI layer. As examples, we study the MR of a metallic film adjacent to either a paramagnet (PM) or a Weiss ferromagnet (FM). 
At low temperatures, we find a striking non-monotonic behavior of the MR as a function of $B$, which we explain in terms of an interplay between the SMR and HMR effects. 
Our model provides a tool to reveal, by MR  measurements, magnetic properties of NM/MI interfaces.

\begin{figure}[th!]
\begin{center}
\includegraphics[width=0.48\textwidth]{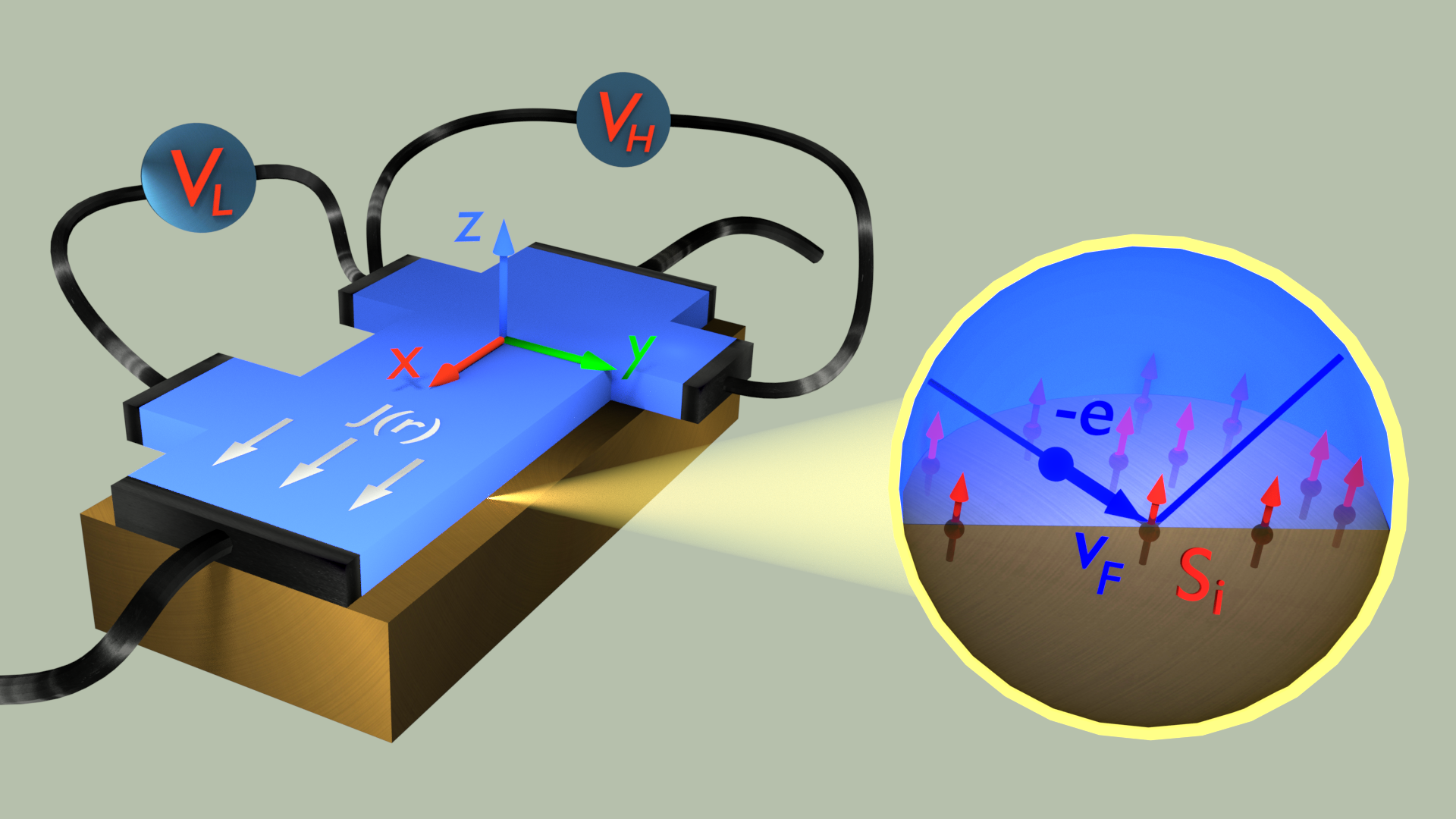} 
\end{center}
\caption{
Sketch of a Hall bar fabricated from a thin metallic film (blue) deposited on the surface of a magnetic insulator (brown). 
The longitudinal ($V_L$) and Hall ($V_H$) voltages are sensitive to variations of the charge current occurring under the influence of the spin-dependent scattering at the interface.
The inset shows the basic process responsible for SMR: 
an electron with charge $-e$
and Fermi velocity $v_F$ moving randomly in the metal 
scatters off the surface of the magnetic insulator and 
interacts with a local moment $\vec{S}_i$.
The SMR corrections are expressed in terms of the interfacial exchange field, the spin-flip rate, and the spin dephasing rate, 
which all depend on the magnetic state of the local moments, and thus, can be controlled by magnetic fields and temperature.
}
\label{FIG0}
\end{figure}

\emph{Model and Method}-
We consider an NM in contact with an MI, as shown in Figure \ref{FIG0}.  We assume both layers to be homogeneous in the $(x,y)$ plane and the NM/MI interface to be located at $z=0$. 
The system Hamiltonian reads $H=H_{\textrm{NM}}+H_{\textrm{MI}}+V_{sd}$,
where
$H_{\textrm{NM}}$ is the Hamiltonian of a disordered metal with SOC and Zeeman field,
$H_{\textrm{MI}}$ is the Heisenberg Hamiltonian in a magnetic field,
and $V_{sd}$ describes the coupling at the NM/MI interface.
We model the interface by an exchange interaction between local moments and itinerant electrons
\begin{equation} 
\label{eq:Vsd}
V_{sd}=-J_{sd}\sum_i\vec{S}^i\cdot\vec{s}(\vec{r}_i),
\end{equation}
where $\vec{S}^i$ is the spin operator of the local moment, $\vec{s}(\vec{r}_i)$ is the operator of the itinerant spin density at position $\vec{r}_i$, and $J_{sd}$ is the \textit{sd}-coupling constant at the NM/MI interface.
We assume that the metal is strongly disordered, such that
the mean-free path $l$ is much smaller than both the thin-film thickness $d_N$
and the spin-relaxation length $\lambda_s$.
For such a \emph{diffusive} motion of the electron in the thin film,
the events of interaction with the local moments located on the surface of the MI
appear as spikes of short duration, randomly distributed along 
the semiclassical trajectory of the electron.
The precise positions of the spikes on the trajectory is clearly unimportant, because the trajectory is sufficiently random.
In this diffusive limit, 
we may allow ourselves to displace 
the local moments in a random fashion
on the scale of $l$ without any consequence for the 
disorder-averaged quantities,
as long as we are interested in the dependence of those quantities on a
 larger scale, set by $\lambda_s$.
Thus, we arrive at considering a fictitious layer of thickness $b$ in which both the itinerant electrons and the local moments coexist, 
with the latter being randomly distributed but maintaining their spin-spin coupling. 
We apply the Born-Markov approximation to $H$ in this $b$-layer, with $V_{sd}$ in Eq.~(\ref{eq:Vsd}) as perturbation. 
Although the thickness $b$ should be kept small ($b\sim l$), we obtain physically meaningful results
by sending first $l\to 0$ in the diffusive limit, and only in a second step $b\to 0$,
going thus through an intermediate stage of the calculation in which $l \ll b\ll \lambda_s$.
This order of taking the limits 
represents a significant simplification in the derivation,
because powerful disorder-averaging techniques devised for homogeneously distributed impurities in the metal 
can be applied here to calculate the spin-relaxation tensor inside the $b$-layer in a local continuum approximation~\footnote{It is important to remark that the coupling in Eq.~(\ref{eq:Vsd}) acts 
more efficiently when the spin $\vec{S}_i$ is embedded in the metal as compared to the case when it is at the surface and interacts only with the tail of the electron wave function appearing in $\vec{s}(\vec{r}_i)$.
We should, therefore, reduce $J_{\textit{sd}}$ in Eq.~(\ref{eq:Vsd}) by a factor $n(z>\lambda_F)/n(z=0)$, 
where $n(\vec{r})$ is the average charge density in the metal
and $\lambda_{F}$ is the Fermi wave length.
However, this suppression factor is expected to be on the order of unity in well-coupled systems, for which the local moments at the surface form bonds with the metal.
We absorb this suppression factor into $J_{sd}$ hereafter.}.

To simplify the magnetic problem, we employ the 
Weiss mean-field theory for $H_{\textrm{MI}}$.
In this approximation, the state of the magnetic system is 
a product state of individual local moments, yielding
\begin{equation}
\langle S^{i}_{\alpha}(t)S^{j}_{\gamma} \rangle=
\delta_{ij}
\langle S^{i}_{\alpha}(t)S^{i}_{\gamma} \rangle
+(1-\delta_{ij})
\langle S^{i}_{\alpha}\rangle \langle S^{j}_{\gamma} \rangle.
\end{equation}
The equilibrium properties of the local moments are determined by the spin expectations $\langle S_{\alpha}\rangle$ and $\langle S^2_{\alpha}\rangle$ ($\alpha=x,y,z$), which depend on $T$ and $B$.  We do not consider here the feedback effect of the itinerant electrons on the local moments;
the latter act merely as a spin bath on the itinerant electrons.

In this approach, we arrive at the usual continuity equation for the non-equilibrium spin bias $\vec{\mu}_s$ 
in the metal (including the $b$-layer) following a standard derivation 
\begin{equation} 
\label{eq:diffusion}
\dot{\mu}_{s}^{\alpha }-\frac{1}{e\nu_F }\partial_i J_{s,i}^{\alpha }-\omega _{L}\epsilon _{\alpha \gamma \kappa }n_{\gamma }\mu
_{\mathrm{s}}^{\kappa }=-\Gamma_{\alpha \kappa}\mu _{s}^{\kappa },
\end{equation}%
where superscript Greek indices denote spin projections ($\alpha = \{x,y,z\}$)
and subscript Latin indices denote current directions ($i=\{x,y,z\}$),
$\vec{n}=\vec{B}/B$ is the unit vector of the $B$-field,
$e$ is the elementary charge ($e>0$),
$\nu_F$ is the density of states per spin species at the Fermi level,
$\epsilon _{\alpha \beta \gamma }$ is the antisymmetric tensor, 
and repeated indices are implicitly summed over.
The spin current $J_{s,i}^{\alpha }$ has units of electrical current,
with $-J_{s,i}^{\alpha}/2e$ giving the amount of spin with polarization $\alpha$ transported in direction $i$ through a unit cross section and per unit of time.
Both the Larmor precession frequency, $\omega _{L}$, and the spin relaxation  tensor, 
$\Gamma_{\alpha \kappa}$ are inhomogeneous in space due to the $b$-layer insertion.
Specifically,  for the geometry in Figure \ref{FIG0}, we have
 \begin{equation} \label{ELF}
    \omega _{L}(z)= \omega_B-\frac{n_{\textit{imp}}^{\textit{2D}}J_{\textit{sd}}}{\hbar}\langle \hat{S}_{\parallel}\rangle \delta_b(z),
\end{equation}
where $\omega_B=\textsl{g}\mu_BB/\hbar$, with $\textsl{g}\approx 2$ being the electron $\textsl{g}$-factor and $\mu_B$ the Bohr magneton,
$n_{\textit{imp}}^{\textit{2D}}$ is the number of local moments per unit area at the MI/NM interface,
$\hat{S}_{\parallel}=\hat{\vec{S}}\cdot\vec{n}$ is the longitudinal spin operator,
and $\delta_b(z)$ equals to $1/b$ in the $b$-region and zero elsewhere.
In the limit $b \to 0$, 
$\delta_b(z)$ tends to the Dirac $\delta$-function.
The second term on the right-hand side in Eq.~(\ref{ELF}) 
describes the interfacial exchange field.
For instance,  this field is particularly well pronounced in Al/EuS,
leading to a directly measurable splitting of the density of states 
in the superconducting regime~\cite{hao1990spin,strambini2017revealing}.

The spin relaxation tensor in Eq. (\ref{eq:diffusion}) reads
\begin{equation}\label{eq:Gamma}
\Gamma_{\alpha \kappa}(z)=  \frac{\delta_{\alpha \kappa}}{\tau_{s}}+\left[\frac{\delta _{\alpha \kappa }}{\tau _{\bot }}+\left( \frac{1}{%
\tau _{\Vert }}-\frac{1}{\tau _{\bot }}\right) n_{\alpha }n_{\kappa }\right]\delta_b(z),
\end{equation}
where $\tau_{s}$ is the spin relaxation time in the NM.
We assume the spin relaxation in the NM to remain isotropic for the experimentally relevant magnetic fields, $B\lesssim 10\;\textrm{T}$.
Indeed, the Pauli paramagnetism has a weak effect
on the SOC-induced spin relaxation at the Fermi level, because the density of states is almost spin-independent,
$\nu^{\uparrow}_F\approx\nu^{\downarrow}_F\equiv \nu_F$, owing to the large Fermi energy of the NM.
In Eq.~(\ref{eq:Gamma}),
$\tau _{\Vert }$  and $\tau _{\bot }$ denote, respectively, the longitudinal  and transverse spin relaxation times per unit length for the itinerant electron in the $b$-region.
In our notations, 
$T_1 = b \tau_{\parallel}$ is the relaxation time of the longitudinal spin component ${S}_{\parallel}=\vec{S}\cdot\vec{n}$, 
and $T_2= b\tau_{\perp}$ is the decoherence time of the transverse spin components $\vec{S}_{\perp}=\vec{n}\times\left(\vec{S}\times\vec{n}\right)$. 
Within the Born-Markov approximation~\cite{slichter2013principles}, we obtain
\begin{eqnarray} \label{tauT}
    \frac{1}{\tau _{\Vert }}&=&\frac{2\pi}{k_BT}  n_{\textit{imp}}^{2D}\nu_F J_{sd}^2 \omega_{m}
 n_{B}\left(\omega_{m}\right)\left[1+n_{B}\left(\omega_{m}\right)\right]\vert\langle \hat{S}_{\Vert
}\rangle\vert, \\  
\frac{1}{\tau _{\bot }} &=&  \frac{1}{2\tau _{\Vert }} + \frac{\pi  }{\hbar } n_{imp}^{2D} \nu_F J^2_{sd}\langle \hat{S}^2_{\Vert
}\rangle,
 \label{tauP}
\end{eqnarray}
where $n_B(\omega)=1/(e^{\hbar\omega/k_BT}-1)$ is the Bose-Einstein distribution function and
$\omega_{m}=\omega_B-\langle \hat{S}_{\Vert}\rangle \sum_{j} J_{ij}/\hbar$,
with $J_{ij}$ being the coupling constant of the Heisenberg ferromagnet.
In deriving Eqs.~(\ref{tauT}) and (\ref{tauP}),
we assumed that the correlator $\left\langle S_\alpha(t)S_{\beta}\right\rangle$ for a spin on the MI surface 
can be approximated by the corresponding correlator for 
a spin deep in the bulk of the MI.
The difference between $1/\tau_{\parallel}$ and $1/\tau_{\perp}$
is entirely due to the ordered magnetic state of the local moments at the interface.

To derive the boundary condition for the NM/MI interface, we
integrate Eq.~(\ref{eq:diffusion}) over $z$ in the 
$b$-layer ($-b<z<0$), assuming 
that $\vec{\mu}_s$ is almost constant and independent of time,
\begin{eqnarray}
-\frac{1}{e\nu_F}\left.J^{\upsilon}_{s,z}\right\vert^{z=0}_{z=-b} 
&=& b\omega_L \epsilon_{\upsilon\gamma\kappa}n_{\gamma} \mu^{\kappa}_s-\left(\frac{b}{\tau_s}+\frac{1}{\tau_{\perp}}\right) \mu^{\upsilon}_s \nonumber\\ &&-\left(\frac{1}{\tau_{\Vert}}-\frac{1}{\tau_{\perp}}\right)n_{\upsilon} (\vec{n}\cdot \vec{\mu}_s).
\label{OURBC_MI}
\end{eqnarray}
Next we take the limit $b\to 0$ and 
write the boundary condition in a customary way~\cite{brataas2001spin,dejene2015control}
\begin{equation}\label{BC_MI}
    -e\vec{J}_{s,z}(0)= G_s\vec{\mu}_s+G_r \vec{n} \times (\vec{n}\times \vec{\mu}_s)+ G_i \vec{n}\times \vec{\mu}_s, 
\end{equation}
where we set $\vec{J}_{s,z}=0$ at $z=-b$, 
because, by construction, 
the electron does not penetrate into the MI beyond the $b$-layer.
The spin dependent conductances read
\begin{align} 
    G_s&=-e^2 \nu_F\frac{1}{\tau_{\parallel}}, \label{eq:Gs}
\\
    G_r&=e^2 \nu_F \left(\frac{1}{\tau_{\perp}}-\frac{1}{\tau_{\parallel}}\right),\label{eq:Gr}
\\
    G_i&=-\frac{e^2}{\hbar} n_{imp}^{2D} \nu_F J_{sd} \langle \hat{S}_{\parallel}\rangle.
\label{eq:Gi}
\end{align}
It is customary to call the complex quantity $G_{\uparrow\downarrow}=G_r+iG_i$ spin-mixing conductance~\cite{brataas2001spin}, 
whereas $G_s$ is sometimes called spin-sink conductance~\cite{dejene2015control}.
We note that $G_s$ originates entirely from spin-flip processes and 
can, therefore, be unambiguously associated with magnon emission and absorption.
In contrast, $G_r$ does not have a physical meaning on its own.
However, the combination $G_r-G_s$ is proportional 
to the spin decoherence rate ($1/\tau_{\perp}$)
of the itinerant electron at the NM/MI interface.
It follows from Eq.~(\ref{tauP}) that a part of $G_r-G_s$ 
is due to spin-flip processes ($1/2\tau_{\parallel}$), 
and hence is identical in nature to 
$G_s$,
whereas the other part is due to spin dephasing.
The purely dephasing contribution is $G_r-\frac{1}{2} G_s$
and it originates from almost elastic spin-scattering processes,
which do not involve a spin exchange with the MI.
Thus, $G_s$ and $G_r-\frac{1}{2} G_s$ correspond to 
different physical processes and have, therefore, distinct 
dependences on $B$ and $T$.  
Finally, $G_i$ is a measure of the interfacial exchange field and it is proportional to the MI magnetization.

\begin{figure*}[th!]
\begin{center}
\includegraphics[width=0.9\textwidth]{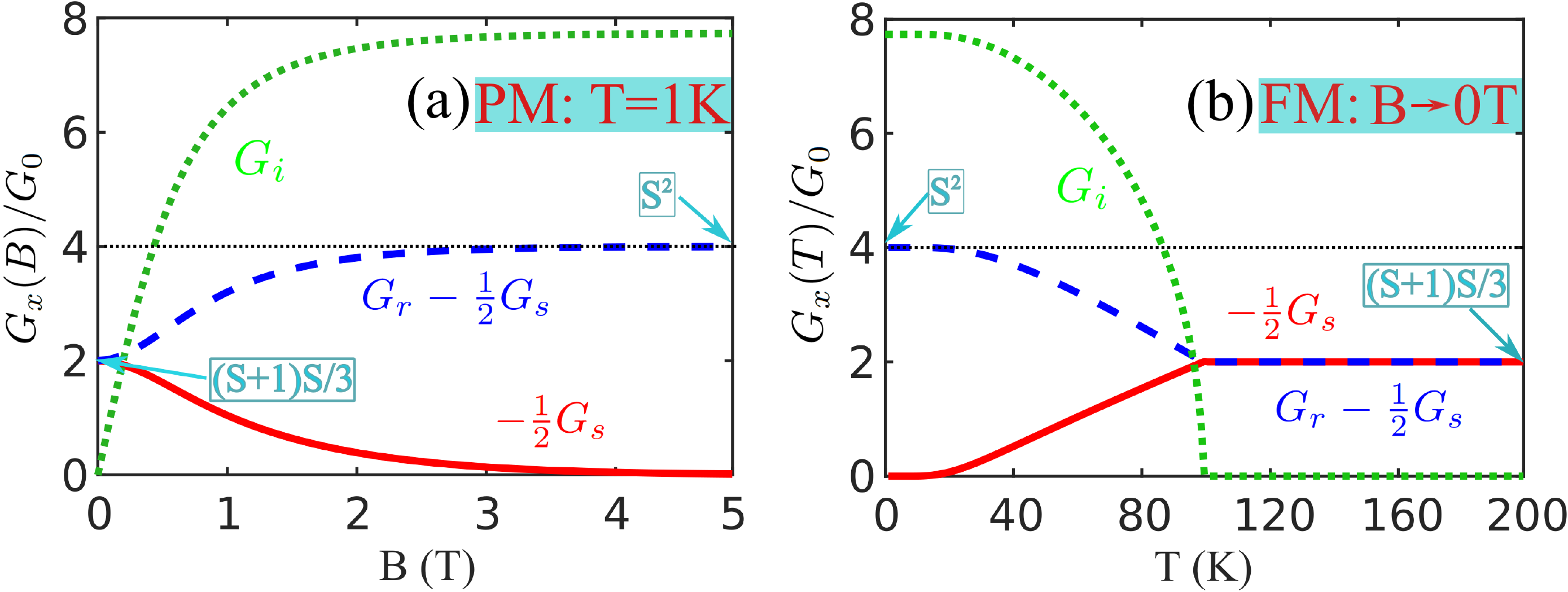}
\end{center}
\caption{
The  spin dependent conductances 
$G_{x}$ ($x=s,r,i$) arranged in combinations of $-\frac{1}{2}G_s$, $G_r-G_s$, and $G_i$ to describe, respectively, the spin flips (magnon emission), the spin dephasing (no spin transfer to MI), and the interfacial exchange field. 
(a) The dependence on $B$ for a PM insulator at $T=1\,\textrm{K}$. 
(b) The dependence on $T$ for an FM insulator with a Curie temperature of $T_{C}=100\,{\textrm{K}}$.
The conductnaces are measured in units of $G_0= \frac{\pi}{\hbar} n_{\textit{imp}}^{\textit{2D}}(e \nu_F J_{sd})^2$.  
The \textit{sd}-coupling constant is $J_{sd}a^{-3}_{c}=0.2$ eV~\cite{wahl2007exchange},
which is parameterized by the lattice constant of the NM, $a_c=0.4\,\textrm{nm}$. Other parameters: $S=2$, 
$ \nu_F J_{sd} \simeq 0.08$, and $n^{\textit{2D}}_{imp}a^2_{c}=0.5$.
}
\label{FIG2}
\end{figure*}

\emph{Results and Discussion}-
We plot the quantities
$G_i$, $G_r-\frac{1}{2}G_s$, and $-\frac{1}{2}G_s$
as functions of $B$ for a PM in Figure~\ref{FIG2}a
and as functions of $T$ for a FM in Figure~\ref{FIG2}b.
In the isotropic regime ($\tau_\parallel=\tau_\perp$),
we have $G_r=G_i=0$, and $G_s=-\frac{2}{3}G_0 S(S+1)$.
In the strongly magnetized regime
($\tau_\parallel\gg\tau_\perp$),
we have
$G_r = G_0 S^2$, $G_i=G_0 S/(\pi\nu_FJ_{\textit{sd}})$, and $G_s\approx 0$.
Here, 
$G_0 = \frac{\pi}{\hbar} n_{\textit{imp}}^{\textit{2D}}(e \nu_F J_{sd})^2$
is a characteristic scale of the spin-dependent conductances.
We estimate a value of $G_0\approx 3.8\times 10^{13}\Omega^{-1}\textrm{m}^{-2}$ 
for a typical $n_{\textit{imp}}^{\textit{2D}}=5\times 10^{18}\textrm{m}^{-2}$ 
and $\nu_F J_{\textit{sd}}=0.1$.
This estimate is compatible with values of spin mixing conductances reported in experiments~\cite{vlietstra2013exchange,dejene2015control,das2019temperature}.

Next we consider a ferrimagnet consisting of two 
species of local moments 
($S^a$ and $S^b$).
In the mean-field approximation, no interference terms
occur between different species
and 
our results above are modified only by selectively weighting each
species by its concentration on the surface ($n_{a}^{\textit{2D}}$ and $n_{b}^{\textit{2D}}$) 
and taking into account its possibly different coupling strength
($J_{\textit{sd}}^{a}$ and $J_{\textit{sd}}^{b}$).
It is possible to obtain a situation in which
the interfacial exchange fields of the two local-moment species 
closely compensate each other, resulting in $G_i\ll G_r$
---a condition which is believed to hold for Pt thin films deposited on
$\textrm{Y}_3\textrm{Fe}_5\textrm{O}_{12}$ (YIG)~\cite{vlietstra2013exchange} and
which would otherwise not be possible in a simple ferromagnet, 
because $G_i$ is the largest spin mixing conductance for
$\nu_FJ_{\textit{sd}}\ll 1$.
The $G_i$-compensation condition for a ferrimagnet, thus, reads 
$n_{a}^{\textit{2D}}J_{\textit{sd}}^{a}S^a - n_{b}^{\textit{2D}}J_{\textit{sd}}^{b}S^b=0$,
which differs from the magnetization compensation condition,
$n_{a}^{\textit{3D}}S^a - n_{b}^{\textit{3D}}S^b=0$,
and allows for the possibility of having a finite magnetization even
when $G_i=0$.
And vice versa, the N\'{e}el order parameter of an antiferromagnet 
can manifest itself as an interfacial exchange field, provided the $G_i$-compensation condition is not fulfilled.
We remark that, for YIG, we have $S^a=S^b=5/2$ and $n_{a}^{\textit{3D}}/n_{b}^{\textit{3D}}=3/2$. 
And for the $\textrm{Pt}/\textrm{YIG}\textrm{-}[001]$ interface, we have $n_{a}^{\textit{2D}}=n_{b}^{\textit{2D}}$ and $J_{\textit{sd}}^{a}\approx J_{\textit{sd}}^{b}$. 
A small difference between $J_{\textit{sd}}^{a}$ and $J_{\textit{sd}}^{b}$ 
may originate from different crystal fields for the $\textrm{Fe}^{3+}$ cation on the tetrahedral ($a$) and octahedral ($b$) sublattice of the garnet.

Despite the fact that YIG has been the material of choice 
in most experimental studies of SMR, recent experiments 
started studying also other 
MIs~\cite{isasa2014spin,isasa2016spin,velez2018spin,lammel2019spin,koichi2019paramagneticSMR,fontcuberta2019role}.
Here, we would like to draw attention to an effect 
due to $G_i$ which, to the best of our knowledge, 
has been overlooked theoretically
and which could appear rather puzzling when observed experimentally.
This effect consists in a negative, linear-in-$B$ magnetoresistance, which arises from an interplay between SMR and HMR featuring a non-local Hanle effect.
And despite the fact that the novel effect is primarily due to $G_i$, we keep $G_r$ and $G_s$ in the expressions below for completeness.

We make use of the boundary condition 
in Eq.~(\ref{BC_MI}) and 
follow closely the derivation of the SMR and HMR effects~\cite{nakayama2013spin,chen2013theory,velez2016hanle,velez2018spin},
obtaining the corrections to the longitudinal ($\rho_L$) 
and transverse $(\rho_T)$ resistivity of the Hall-bar setup
in Figure~\ref{FIG0}
\begin{eqnarray}
\rho _{L} &\simeq& \rho_D+\Delta
\rho _{0}+\Delta \rho_{1}\left(1-n_{y}^{2}\right),\nonumber\\ 
\rho _{T}&\simeq& -\rho_D\omega_c\tau n_z
+\Delta \rho _{1}n_x n_y+\Delta \rho _{2}n_z,
\label{SLT}
\end{eqnarray}
where $\rho_D$ is the Drude resistivity and
$\omega_c\tau$ is the Hall angle,
with $\omega_c=eB/mc$ being the cyclotron frequency and
$\tau$ being the momentum relaxation time.
The combined $\textrm{SMR+HMR}$ resistivity corrections 
read~\cite{velez2018spin}
\begin{eqnarray}
\Delta \rho _{0}&=&\theta_{\textit{SH}}^{2}\rho_D[2-\mathcal{R}(G_s,\lambda_s)],
\nonumber\\ 
\Delta \rho _{1}&=&\theta_{\textit{SH}}^{2}\rho_D 
\left\{\mathcal{R}(G_s,\lambda_s)-\mathrm{Re} 
\left[ \mathcal{R}(G_s-G_{\uparrow\downarrow},\Lambda)\right]\right\},\quad
\nonumber\\
\Delta \rho _{2}&=&\theta_{\textit{SH}}^{2}\rho_D \mathrm{Im} 
\left[ \mathcal{R}(G_s-G_{\uparrow\downarrow},\Lambda)\right],
\label{DELTA012}
\end{eqnarray}
where $\theta_{\textit{SH}}$ is the spin Hall angle, 
$\lambda_s=\sqrt{\mathcal{D}\tau_s}$, with $\mathcal{D}=1/2e^2\nu_F\rho_D$ being the diffusion constant,
$G_{\uparrow\downarrow}=G_r+iG_i$ is the complex spin mixing conductance, $1/\Lambda=\sqrt{1/\lambda _{s}^{2}+i\omega_B/\mathcal{D}}$,
and the auxiliary function $\mathcal{R}(\mathcal{G},\ell)$
is defined as
\begin{equation}
\mathcal{R}(\mathcal{G},\ell)=
\frac{2\ell}{d_{N}}
\tanh\left(\frac{d_{N}}{2\ell}\right)
\frac{1-\rho_D\mathcal{G}\ell 
\coth\left(\frac{d_{N}}{2\ell}\right)}
{1-2\rho_D\mathcal{G}\ell
\coth\left(\frac{d_{N}}{\ell}\right)}.
\label{gammaell}
\end{equation}
For $\Lambda=\lambda_s$, we recover the
SMR corrections~\cite{nakayama2013spin,chen2013theory},
whereas for $G_{s}=G_{r}=G_{i}=0$, we recover the HMR corrections~\cite{dyakonov2007magnetoresistance,velez2016hanle}.
We remark that, in general, 
it is important to take into account
$G_s$~\cite{dejene2015control,velez2018spin}, which is 
not negligible in the paramagnetic regime, see Figure~\ref{FIG2}.
The corrections in Eq.~(\ref{DELTA012}) were used in Ref.~\citenum{velez2018spin}
to explain the unusual behavior of SMR in 
$\textrm{Pt}\textrm{/}\textrm{La}\textrm{Co}\textrm{O}_3$ in the high-temperature limit.

\begin{figure*}[th!]
\begin{center}
\vspace{0pt}\includegraphics[width=0.85\textwidth]{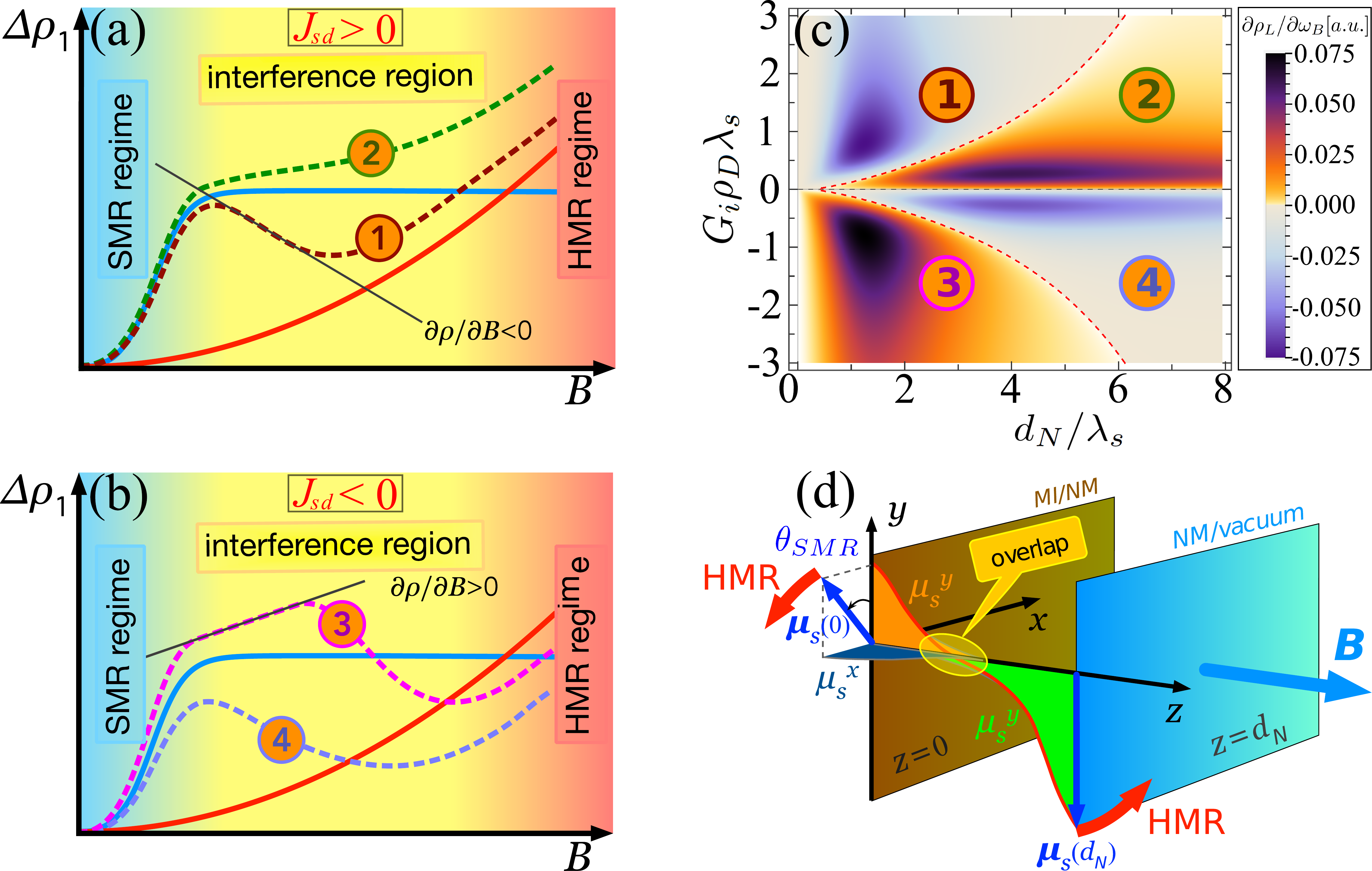}\par
\end{center}
\caption{(a) Interplay between SMR and HMR effects for $J_{\textit{sd}}>0$ and MI in the paramagnetic regime.
The blue and red solid lines show, respectively, the SMR and HMR effects in the absence of one another.
Between the SMR regime (small $B$) and the HMR regime (large $B$) an intermediate ``interference'' region occurs, where anomalous behavior, marked by the straight black line, is possible due to a non-local Hanle effect and its interplay with SMR.
The dashed line \ding{192} shows the qualitative behavior of $\Delta\rho_1(B)$ for the non-local interference,
whereas line \ding{193} shows it for the local interference.
(b) Same as in (a), but for $J_{sd}<0$ and with lines \ding{194} and \ding{195} corresponding, respectively, to the non-local and local regimes of interference. 
(c) Separation of the parameter space $(d_N, G_i)$ into different regimes of interference. 
The red dashed line shows the critical value of $G_i$ in Eq.~(\ref{eqGicrit}) as a function of $d_N$.
The regions \ding{192}-\ding{195} correspond to the four kinds of behavior shown in (a) and (b).
The color code shows the sign of $\partial \rho_L/\partial \omega_B$ at constant $G_i$ and $\omega_B\to 0$. 
(d) Sketch of the spin accumulation at the thin-film interfaces 
as created by the SHE and altered by the SMR effect.
The SMR effect suppresses the spin density at NM/MI interface and
rotates it by a finite angle, $\theta_{\textrm{SMR}}$, about the magnetization direction.
In the absence of overlap between the two spin accumulations (orange and green), the Hanle effect acts locally at each interface and alters the spin accumulation in an expected manner, quite similarly to the SMR effect, see text.
The overlap between the two spin accumulations makes it possible for the Hanle effect from the NM/vacuum interface to affect significantly the spin accumulation at the NM/MI interface, especially when the latter is strongly suppressed due to SMR. 
With applying a $B$ field, the component $\mu_{s}^{y}(z=0)$ can exhibit an increase instead of the decrease which one could na\"{i}vely expect from the local Hanle effect.
}
\label{FIGsketch}
\end{figure*}

For a PM or FM at sufficiently low temperatures,
the scale to reach saturation represents only a relatively small portion of the experimentally accessible $B$-field range.
The SMR effect develops quickly with increasing $B$ and saturates
as shown by the blue solid line in Figure~\ref{FIGsketch}a.
The SMR effect is dominated by $G_i$ for
\begin{equation}
G_r-G_s\ll  G_i^2 \lambda_s\rho_D \coth{\left({d_N}/{\lambda_s}\right)},
\end{equation}
which requires that $n_{\textit{imp}}^{\textit{2D}}\gg (\hbar/e^2\rho_D\lambda_s)\tanh{\left(d_N/\lambda_s\right)}$.
At the same time, the HMR effect develops gradually and
becomes relevant only for large $B$
as shown by the red solid line in Figure~\ref{FIGsketch}a.
In experiment, the HMR effect is well pronounced at relatively large magnetic fields, $B\lesssim 10\,\textrm{T}$~\cite{velez2016hanle}.
In the region of intermediate $B$, denoted 
as ``interference region'' in Figure~\ref{FIGsketch}a, 
the interplay between the
SMR and HMR effects can lead to negative differential MR 
($\partial \rho_L/\partial B <0$).
This behavior would not be so surprising if it occurred solely when
$G_i$ and $\omega_B$ had opposite signs.
Indeed, $G_i$ is a measure of the interfacial exchange field,
which is a singular field created at the NM/MI interface by the
\textit{sd} coupling in Eq.~(\ref{eq:Vsd}).
The signs of $G_i$ and $\omega_B$ are equal to each other for
$J_{\textit{sd}}>0$ and opposite for
$J_{\textit{sd}}<0$.
For electrons diffusing over a characteristic length scale $\ell\sim\min(d_N,\lambda_s)$,
the interfacial exchange field can be smeared near the interface over $\ell$ and superimposed onto $\omega_B$, obtaining an average Larmor frequency $\omega_L=\omega_B+G_i/e^2\nu_F\ell$.
One could na\"{i}vely expect that the HMR effect, which is
proportional to $\omega_B^2$ for all experimentally relevant $B$-field values, 
to become proportional to $\omega_L^2=\left(\omega_B+G_i/e^2\nu_F\ell\right)^2$,
generating, thus, after squaring a cross term proportional to 
$\omega_BG_i$.
For $J_{\textit{sd}}<0$, this term would then naturally lead to a negative MR.
Quite surprisingly, we find a negative MR even for $J_{\textit{sd}}>0$, provided $G_i$ exceeds a certain critical value. 

To investigate the origin of the anomalous 
behavior of the MR,
we expand $\Delta\rho_1$ in Eq.~(\ref{DELTA012}) in powers of 
$\omega_B$ at constant $G_i$ and set, for simplicity, $G_r=G_s=0$.
The coefficient in front of the linear-in-$\omega_B$ term 
changes sign at the critical value of $G_i$ given by
\begin{equation}
G_{i,c}^2=
\frac{\sinh^2(d_N/2\lambda_s)}{2\lambda_s^2\rho_D^2\cosh(d_N/\lambda_s)}
\left[\frac{\lambda_s}{d_N}\sinh\left(\frac{d_N}{\lambda_s}\right)-1\right].\label{eqGicrit}
\end{equation}
We find several qualitatively different behaviors of the MR,
illustrated 
by the dashed lines in Figure~\ref{FIGsketch}a-b.
The lines \ding{192}--\ding{195} correspond to the regions 
in the parameter space $(d_N,G_i)$ shown in Figure~\ref{FIGsketch}c,
obtained by plotting the magnitude of the linear-in-$\omega_B$ term. 
The critical value in Eq.~(\ref{eqGicrit}) as a function of $d_N$ 
is shown in Figure~\ref{FIGsketch}c by the red dashed line.
Thus, for $J_{\textit{sd}}>0$, 
the anomalous behavior manifests itself in
a segment of negative MR on line \ding{192}, marked by the black straight line in Figure~\ref{FIGsketch}a.
The dependence shown by line \ding{193} is consistent with the 
physical picture given above, in which the Zeeman and exchange fields
can be superimposed locally with one another, giving rise to a
shifted-to-the-left parabolic $B$-field dependence for $\Delta\rho_1(B)$, 
on top of the fully developed SMR gap.
The dependence shown by line \ding{192} cannot be understood 
in terms of a local interplay between the SMR and HMR effects
occurring at the NM/MI interface.
We remark that no anomalous behavior occurs in a semi-infinite space, 
at one interface.
Therefore, it is essential to involve in the explanation the NM/vacuum interface, 
which has a spin accumulation oriented predominantly opposite to, 
but not strictly anti-aligned with the spin accumulation at the NM/MI interface.

We illustrate the spin accumulations occurring in the SMR effect at both interfaces in Figure~\ref{FIGsketch}d.
Since the SMR effect suppresses significantly 
the spin accumulation $\vec{\mu}_{s}(0)$ 
at the NM/MI interface, the Hanle effect occurring 
near that interface induces a rather small change of spin accumulation, which represents mainly a rotation 
about the $B$-field axis, such as $\delta\vec{\mu}_{s}(0)\propto \omega_B\left[\vec{n}\times\vec{\mu}_{s}(0)\right]$.
In contrast, the Hanle effect occurring 
near the NM/vacuum interface induces, in the same fashion, a relatively larger change of spin accumulation,
$\delta\vec{\mu}_{s}(d_N)\propto \omega_B\left[\vec{n}\times\vec{\mu}_{s}(d_N)\right]$.
By means of diffusion or, in other words, when the film is so thin that the spin accumulations of both interfaces overlap with each other (see orange and green parts of $\mu_s^y$ in Figure~\ref{FIGsketch}d),
a non-local interplay between SMR and HMR effects takes place.
In particular, for a magnetic field along $z$ as shown in Figure~\ref{FIGsketch}d,
the Hanle effect at the NM/vacuum interface 
brings in a $\mu_s^x$ component generated from a $\mu_s^y$ component
of opposite sign (green part of $\mu_s^y$).
After diffusing across the thin film thickness, the $\mu_s^x$ component is converted back into a $\mu_s^y$ component at the NM/MI interface, due to the interfacial exchange field.
The longitudinal resistivity correction is governed by
the change in the $y$-component of the spin bias across the sample~\cite{chen2013theory,velez2016hanle},
$\Delta\rho_L\propto \mu_s^{y}(d_N)-\mu_s^{y}(0)$.
A negative MR is obtained when the difference 
$\mu_s^{y}(0)-\mu_s^{y}(d_N)$ grows with applying a magnetic field,
\emph{i.e.}\ when the spin bias across the sample increases with $B$.
This usual behavior is obtained also for the quantity $\mu_s^{y}(0)$ alone, although we find that the difference $\mu_s^{y}(0)-\mu_s^{y}(d_N)$
begins to increase with $B$ at a smaller 
critical $G_i$ than the value at which $\mu_s^{y}(0)$ begins to increase.
Nevertheless, the physical picture leading to such a striking  effect
is common to both quantities:
The $\mu_s^x$ component generated from 
a large negative spin accumulation at the 
NM/vacuum interface is converted into 
a $\mu_s^y$ component at the NM/MI interface due to $G_i$,
obtaining a non-local contribution
$\delta \mu_s^{y}(0)\propto -\omega_B G_i e^{-d_N/\lambda_s}\mu_s^y(d_N)$.
This non-local contribution competes with 
the one generated locally by the Hanle effect at the NM/MI interface,
$\delta \mu_s^{y}(0)\propto \omega_B \mu_s^x(0)$.
Notably, $\mu_s^x(0)$ is suppressed for large $G_i$ as $\propto 1/G_i$,
which makes the correction generated locally small.
From the balance of the local and non-local corrections to 
$\mu_s^{y}(0)$,
we recover the exponential dependence of the critical 
$G_i$ in Eq.~(\ref{eqGicrit}) for large $d_N\gg\lambda_s$, 
namely $G_{i,c}\propto e^{-d_N/2\lambda_s}$.
Thus, we conclude that the transition from positive to negative MR for $J_{\textit{sd}}>0$
occurs when the non-local interplay between HMR and SMR 
dominates over the local one.

In the case of $J_{\textit{sd}}<0$, see Figure~\ref{FIGsketch}b,
a negative MR is not unusual, since the Zeeman and exchange fields have opposite signs and can, thus, compensate each other to some extent. 
In this case, one would expect a shifted-to-the-right parabolic $B$-field dependence for $\Delta\rho_1(B)$, on top of the fully developed SMR gap.
This expectation is, indeed, met when the magnitude of $G_i$ is smaller than the critical value in Eq.~(\ref{eqGicrit}), see line \ding{195} in Figure~\ref{FIGsketch}b.
As for line \ding{194}, which corresponds to a large negative $G_i$ ($G_i<G_{i,c}<0$),
its behavior resembles qualitatively that of line \ding{195},
and can not be reliably identified in the absence 
of the reference curves showing the pure SMR and pure HMR separately.
Nevertheless, the anomalous behavior originating from the
non-local interplay between SMR and HMR 
consists here in having a positive slope 
in the beginning of the interference region, as marked by the straight black line in Figure~\ref{FIGsketch}b.

With the help of our theoretical model we explore further  several examples that illustrate the non-monotonic behavior of the MR in a realistic system and show how it evolves with temperature. Specifically, we assume that the MI can be described as a Weiss ferromagnetic insulator.  It  exhibits a spontaneous finite average magnetization, $\langle \hat{S}_{\parallel }\rangle$, at temperatures below   the  Curie-Weiss temperature $T_{c}$. The $B$- and $T$- dependence of $\langle \hat{S}_{\parallel }\rangle$ is obtained by solving  the transcendental equation, 
\begin{equation}
\label{eq:Weiss}
    \langle \hat{S}_{\Vert } \rangle=-S\mathcal{B}_S\left[ S(\hbar\omega_B-6\langle \hat{S}_{\Vert } \rangle J_{m})/T\right],
\end{equation}
where $\mathcal{B}_S(X)$ is Brillouin function and $J_{m}$ is the coupling constant between nearest neighbours in the Heisenberg model. This expression also describes  a  PM insulator  after setting $J_m=0$.

\begin{figure*}[th]
\begin{center}
\includegraphics[width=0.98\textwidth]{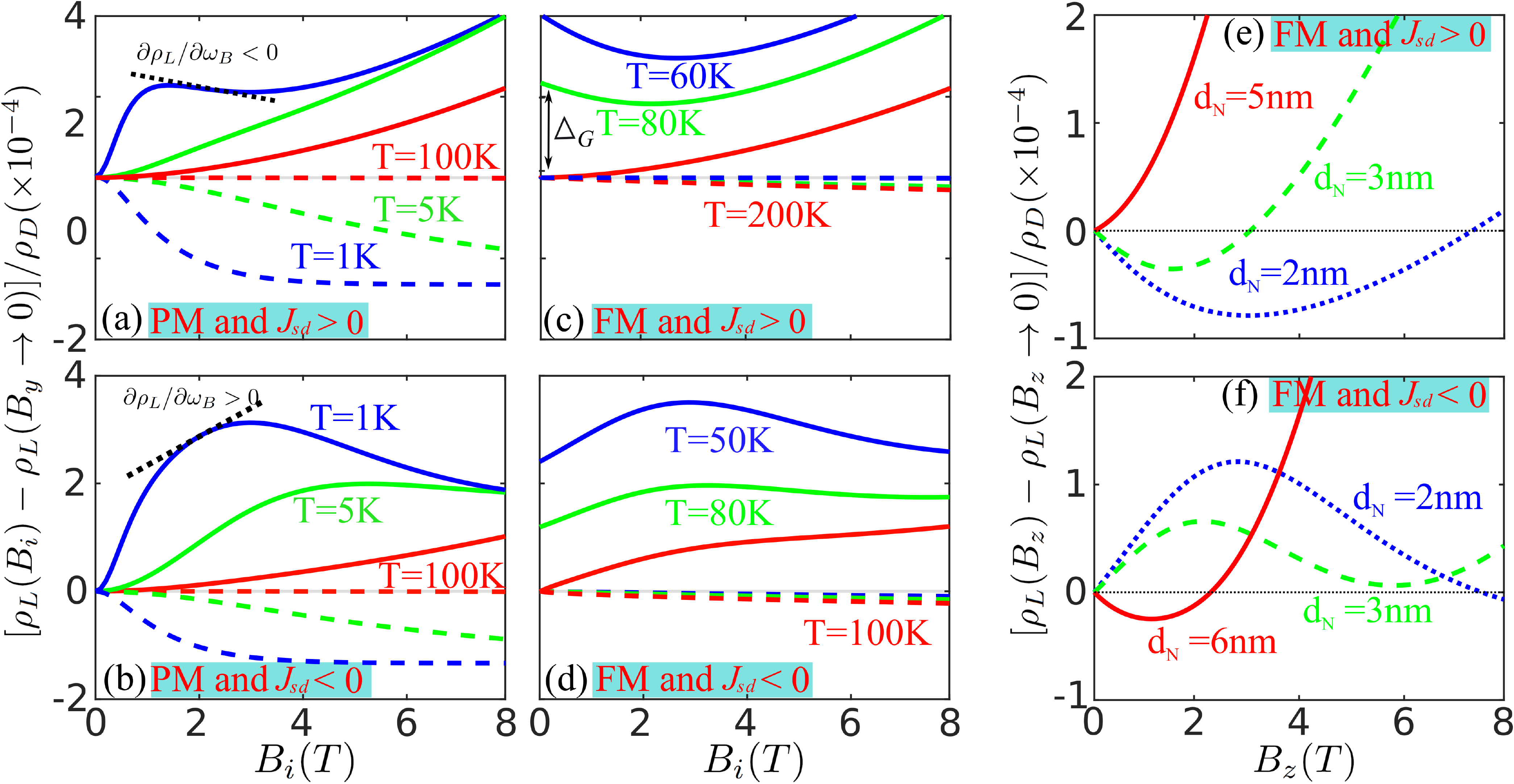} %
\end{center}
\caption{(a-d) Relative longitudinal resistivity, $[\rho _{L}(B_i)-\rho _{L}(B_y\rightarrow 0)]/\rho _{D}$, as a function of magnetic field $B$ applied along main directions ($B\equiv B_i$ with $i=x,y,z$). Solid lines show the case $B\equiv B_z$ (or $B\equiv B_x$, which is equivalent), whereas dashed lines shows the case $B\equiv B_y$.
Panels (a,b) correspond to a PM insulator, whereas panels (c,d) to a FM insulator with a Curie temperature of $T_C=100\,\textrm{K}$.
Panels (a,c) and (b,d) correspond, respectively, to positive and negative  \textit{sd}-coupling constant, $J_{sd}a^{-3}_{c}=\pm 0.1$ eV~\cite{wahl2007exchange,myers2005antiferromagnetic}, respectively. The NM thickness is chosen as $d_N=2$ nm and the different curves correspond to different temperatures.  
(e-f) Different to above, relative longitudinal resistivity, $[\rho _{L}(B_z)-\rho _{L}(B_z\rightarrow 0)]/\rho _{D}$, as a function of magnetic field $B$ applied in $z$-direction,
for a FM insulator which couples to NM with a coupling constant: (e) $J_{sd}a^{-3}_{c}=+0.1$ eV and (f) $J_{sd}a^{-3}_{c}=-0.1$ eV. 
The temperature is chosen as $T=10\,\textrm{K}$ and the different curves correspond to different thickness. In all panels we have chosen the following values of other parameters: $\theta _{SH}=0.1$, $\lambda_{s}=3.0\,\textrm{nm}$, 
$\rho_{D}=1.0\times 10^{-6}\;\Omega\;\!\textrm{m}$, $n^{2D}_{imp}a^2_{c}=0.5$, $S=2$, 
$\vert \nu_F J_{sd}\vert \simeq 0.04$, 
and $a_c=0.4\,\textrm{nm}$.}
\label{FIG1}
\end{figure*}

We compute the longitudinal resistivity from Eqs. (\ref{SLT}-\ref{gammaell}). The spin-dependent conductances, Eqs. (\ref{eq:Gs}-\ref{eq:Gi}), are determined from  the  relaxation times in Eqs. (\ref{tauT}-\ref{tauP}), which can be obtained after substitution of magnetization $ \langle \hat{S}_{\Vert } \rangle$ from  Eq. (\ref{eq:Weiss}) and spin-spin correlation function $\langle S^2_{\parallel} \rangle$ from relation $\langle S^2_{\parallel} \rangle=S(S+1)+\coth[(\hbar\omega_B-6\langle \hat{S}_{\Vert } \rangle J_{m})/2T]\left\langle S_{\parallel} \right\rangle$. 
Figure \ref{FIG1} summarizes our  results for a PM and a FM insulators. The dashed lines in  Figure \ref{FIG1}a-d correspond to a field applied in $y$-direction, whereas the solid lines to a field in $z$-direction. It is in  the latter situation that the  predicted anomalous  behavior becomes evident.

As one might anticipate, the non-monotonic behavior is best pronounced at low temperatures for which the spin-dependent conductances $G_{s}$, $G_{r}$, and $G_{i}$ saturate after applying a relatively small magnetic field. 
We have chosen the parameters  such that the solid curves in Figure \ref{FIG1}a-d correspond to the predicted anomalous behaviors  \ding{192} and \ding{194} in Figure \ref{FIGsketch}a-b.
In the PM case, Figure \ref{FIG1}a-b, the anomalous differential MR starts at finite fields when  $G_i$ is sufficiently large, {\it cf.} Figure \ref{FIG2}a.    In contrast,  in the  FM case,  $G_i$ is large enough even at small fields due to the spontaneous magnetization, and the anomalous behaviors are already seen for  $B\to 0$  and  over a larger range of temperatures below $T_c$, see Figure \ref{FIG1}c-d. In the FM case one also obtains the SMR gap,  defined as 
$\Delta_G := \rho _{L}(B\equiv B_{z}\rightarrow 0)-\rho _{L}(B\equiv B_{y}\rightarrow 0)$. 

In Figure \ref{FIG1}e-f, we show $\rho_L(B)$ in the FM case  
for a field in $z$-direction and different values of the NM thickness, $d_N$. 
In accordance to Eq.~\ref{eqGicrit}, by changing $d_N$ one tunes  the critical value of $G_i$ and hence the 
behavior of the MR changes.   For the chosen parameters in Figure \ref{FIG1}e-f,  the thickest film  exhibits the normal behavior, see red solid lines in Figure \ref{FIG1}e-f, whereas thinner films show the anomaly in the  MR, blue and green dashed lines. Thus, our modelling shows that the anomalous behavior is expected to be observed in MIs with sufficiently large values of $G_i$. This can be achieved for example  in insulating FMs with large local moments, as for example in EuS or EuO~\cite{muller2009thickness,strambini2017revealing,wei2016strong}.

\emph{Conclusions}-
We have presented a theory of the SMR effect from a microscopic perspective, in which SMR relates to the microscopic processes of spin relaxation at the NM/MI interface. 
Our theory covers a wide range of MIs and can be used to investigate the effect of a magnetic field and temperature on MR in NM/MI Hall-bar setups and beyond. We found a non-local interplay between SMR and HMR which gives rise to a negative linear-in-magnetic-field MR.
Our theory provides a useful tool for understanding present and future experiments and it has the potential to evolve into a full-fledged technique to measure the magnetic properties of the NM/MI interfaces, focusing exclusively on probing the very surface of the MI.

\emph{Acknowledgement}-
This work was supported by Spanish Ministerio de Economia y Competitividad (MINECO) through the Projects No. FIS2014-55987-P 
and FIS2017-82804-P, and EU’s Horizon 2020 research and innovation program under Grant Agreement No. 800923 (SUPERTED). We thank Felix Casanova, Saul Velez, and Yuan Zhang for useful discussions.


\begin{thebibliography}{46}%
\makeatletter
\providecommand \@ifxundefined [1]{%
 \@ifx{#1\undefined}
}%
\providecommand \@ifnum [1]{%
 \ifnum #1\expandafter \@firstoftwo
 \else \expandafter \@secondoftwo
 \fi
}%
\providecommand \@ifx [1]{%
 \ifx #1\expandafter \@firstoftwo
 \else \expandafter \@secondoftwo
 \fi
}%
\providecommand \natexlab [1]{#1}%
\providecommand \enquote  [1]{``#1''}%
\providecommand \bibnamefont  [1]{#1}%
\providecommand \bibfnamefont [1]{#1}%
\providecommand \citenamefont [1]{#1}%
\providecommand \href@noop [0]{\@secondoftwo}%
\providecommand \href [0]{\begingroup \@sanitize@url \@href}%
\providecommand \@href[1]{\@@startlink{#1}\@@href}%
\providecommand \@@href[1]{\endgroup#1\@@endlink}%
\providecommand \@sanitize@url [0]{\catcode `\\12\catcode `\$12\catcode
  `\&12\catcode `\#12\catcode `\^12\catcode `\_12\catcode `\%12\relax}%
\providecommand \@@startlink[1]{}%
\providecommand \@@endlink[0]{}%
\providecommand \url  [0]{\begingroup\@sanitize@url \@url }%
\providecommand \@url [1]{\endgroup\@href {#1}{\urlprefix }}%
\providecommand \urlprefix  [0]{URL }%
\providecommand \Eprint [0]{\href }%
\providecommand \doibase [0]{http://dx.doi.org/}%
\providecommand \selectlanguage [0]{\@gobble}%
\providecommand \bibinfo  [0]{\@secondoftwo}%
\providecommand \bibfield  [0]{\@secondoftwo}%
\providecommand \translation [1]{[#1]}%
\providecommand \BibitemOpen [0]{}%
\providecommand \bibitemStop [0]{}%
\providecommand \bibitemNoStop [0]{.\EOS\space}%
\providecommand \EOS [0]{\spacefactor3000\relax}%
\providecommand \BibitemShut  [1]{\csname bibitem#1\endcsname}%
\let\auto@bib@innerbib\@empty
\bibitem [{\citenamefont {Dyakonov}\ and\ \citenamefont
  {Perel}(1971{\natexlab{a}})}]{DyakonovPerel1971JETPL}%
  \BibitemOpen
  \bibfield  {author} {\bibinfo {author} {\bibfnamefont {M.~I.}\ \bibnamefont
  {Dyakonov}}\ and\ \bibinfo {author} {\bibfnamefont {V.~I.}\ \bibnamefont
  {Perel}},\ }\href@noop {} {\bibfield  {journal} {\bibinfo  {journal} {Pis'ma
  Zh. Eksp. Teor. Fiz.}\ }\textbf {\bibinfo {volume} {13}},\ \bibinfo {pages}
  {657} (\bibinfo {year} {1971}{\natexlab{a}})}\BibitemShut {NoStop}%
\bibitem [{\citenamefont {Dyakonov}\ and\ \citenamefont
  {Perel}(1971{\natexlab{b}})}]{DyakonovPerel1971PLA}%
  \BibitemOpen
  \bibfield  {author} {\bibinfo {author} {\bibfnamefont {M.~I.}\ \bibnamefont
  {Dyakonov}}\ and\ \bibinfo {author} {\bibfnamefont {V.~I.}\ \bibnamefont
  {Perel}},\ }\href {\doibase 10.1016/0375-9601(71)90196-4} {\bibfield
  {journal} {\bibinfo  {journal} {Phys. Lett. A}\ }\textbf {\bibinfo {volume}
  {35}},\ \bibinfo {pages} {459} (\bibinfo {year}
  {1971}{\natexlab{b}})}\BibitemShut {NoStop}%
\bibitem [{\citenamefont {Hirsch}(1999)}]{hirsch1999spin}%
  \BibitemOpen
  \bibfield  {author} {\bibinfo {author} {\bibfnamefont {J.}~\bibnamefont
  {Hirsch}},\ }\href@noop {} {\bibfield  {journal} {\bibinfo  {journal}
  {Physical Review Letters}\ }\textbf {\bibinfo {volume} {83}},\ \bibinfo
  {pages} {1834} (\bibinfo {year} {1999})}\BibitemShut {NoStop}%
\bibitem [{\citenamefont {Sinova}\ \emph {et~al.}(2004)\citenamefont {Sinova},
  \citenamefont {Culcer}, \citenamefont {Niu}, \citenamefont {Sinitsyn},
  \citenamefont {Jungwirth},\ and\ \citenamefont
  {MacDonald}}]{sinova2004universal}%
  \BibitemOpen
  \bibfield  {author} {\bibinfo {author} {\bibfnamefont {J.}~\bibnamefont
  {Sinova}}, \bibinfo {author} {\bibfnamefont {D.}~\bibnamefont {Culcer}},
  \bibinfo {author} {\bibfnamefont {Q.}~\bibnamefont {Niu}}, \bibinfo {author}
  {\bibfnamefont {N.}~\bibnamefont {Sinitsyn}}, \bibinfo {author}
  {\bibfnamefont {T.}~\bibnamefont {Jungwirth}}, \ and\ \bibinfo {author}
  {\bibfnamefont {A.}~\bibnamefont {MacDonald}},\ }\href@noop {} {\bibfield
  {journal} {\bibinfo  {journal} {Physical Review Letters}\ }\textbf {\bibinfo
  {volume} {92}},\ \bibinfo {pages} {126603} (\bibinfo {year}
  {2004})}\BibitemShut {NoStop}%
\bibitem [{\citenamefont {Valenzuela}\ and\ \citenamefont
  {Tinkham}(2006)}]{valenzuela2006direct}%
  \BibitemOpen
  \bibfield  {author} {\bibinfo {author} {\bibfnamefont {S.~O.}\ \bibnamefont
  {Valenzuela}}\ and\ \bibinfo {author} {\bibfnamefont {M.}~\bibnamefont
  {Tinkham}},\ }\href@noop {} {\bibfield  {journal} {\bibinfo  {journal}
  {Nature}\ }\textbf {\bibinfo {volume} {442}},\ \bibinfo {pages} {176}
  (\bibinfo {year} {2006})}\BibitemShut {NoStop}%
\bibitem [{\citenamefont {Kimura}\ \emph {et~al.}(2007)\citenamefont {Kimura},
  \citenamefont {Otani}, \citenamefont {Sato}, \citenamefont {Takahashi},\ and\
  \citenamefont {Maekawa}}]{kimura2007room}%
  \BibitemOpen
  \bibfield  {author} {\bibinfo {author} {\bibfnamefont {T.}~\bibnamefont
  {Kimura}}, \bibinfo {author} {\bibfnamefont {Y.}~\bibnamefont {Otani}},
  \bibinfo {author} {\bibfnamefont {T.}~\bibnamefont {Sato}}, \bibinfo {author}
  {\bibfnamefont {S.}~\bibnamefont {Takahashi}}, \ and\ \bibinfo {author}
  {\bibfnamefont {S.}~\bibnamefont {Maekawa}},\ }\href@noop {} {\bibfield
  {journal} {\bibinfo  {journal} {Physical Review Letters}\ }\textbf {\bibinfo
  {volume} {98}},\ \bibinfo {pages} {156601} (\bibinfo {year}
  {2007})}\BibitemShut {NoStop}%
\bibitem [{\citenamefont {Kato}\ \emph {et~al.}(2004)\citenamefont {Kato},
  \citenamefont {Myers}, \citenamefont {Gossard},\ and\ \citenamefont
  {Awschalom}}]{kato2004observation}%
  \BibitemOpen
  \bibfield  {author} {\bibinfo {author} {\bibfnamefont {Y.~K.}\ \bibnamefont
  {Kato}}, \bibinfo {author} {\bibfnamefont {R.~C.}\ \bibnamefont {Myers}},
  \bibinfo {author} {\bibfnamefont {A.~C.}\ \bibnamefont {Gossard}}, \ and\
  \bibinfo {author} {\bibfnamefont {D.~D.}\ \bibnamefont {Awschalom}},\
  }\href@noop {} {\bibfield  {journal} {\bibinfo  {journal} {Science}\ }\textbf
  {\bibinfo {volume} {306}},\ \bibinfo {pages} {1910} (\bibinfo {year}
  {2004})}\BibitemShut {NoStop}%
\bibitem [{\citenamefont {Sih}\ \emph {et~al.}(2005)\citenamefont {Sih},
  \citenamefont {Myers}, \citenamefont {Kato}, \citenamefont {Lau},
  \citenamefont {Gossard},\ and\ \citenamefont {Awschalom}}]{sih2005spatial}%
  \BibitemOpen
  \bibfield  {author} {\bibinfo {author} {\bibfnamefont {V.}~\bibnamefont
  {Sih}}, \bibinfo {author} {\bibfnamefont {R.}~\bibnamefont {Myers}}, \bibinfo
  {author} {\bibfnamefont {Y.}~\bibnamefont {Kato}}, \bibinfo {author}
  {\bibfnamefont {W.}~\bibnamefont {Lau}}, \bibinfo {author} {\bibfnamefont
  {A.}~\bibnamefont {Gossard}}, \ and\ \bibinfo {author} {\bibfnamefont
  {D.}~\bibnamefont {Awschalom}},\ }\href@noop {} {\bibfield  {journal}
  {\bibinfo  {journal} {Nature Physics}\ }\textbf {\bibinfo {volume} {1}},\
  \bibinfo {pages} {31} (\bibinfo {year} {2005})}\BibitemShut {NoStop}%
\bibitem [{\citenamefont {Wunderlich}\ \emph {et~al.}(2005)\citenamefont
  {Wunderlich}, \citenamefont {Kaestner}, \citenamefont {Sinova},\ and\
  \citenamefont {Jungwirth}}]{wunderlich2005experimental}%
  \BibitemOpen
  \bibfield  {author} {\bibinfo {author} {\bibfnamefont {J.}~\bibnamefont
  {Wunderlich}}, \bibinfo {author} {\bibfnamefont {B.}~\bibnamefont
  {Kaestner}}, \bibinfo {author} {\bibfnamefont {J.}~\bibnamefont {Sinova}}, \
  and\ \bibinfo {author} {\bibfnamefont {T.}~\bibnamefont {Jungwirth}},\
  }\href@noop {} {\bibfield  {journal} {\bibinfo  {journal} {Physical Review
  Letters}\ }\textbf {\bibinfo {volume} {94}},\ \bibinfo {pages} {047204}
  (\bibinfo {year} {2005})}\BibitemShut {NoStop}%
\bibitem [{\citenamefont {Zhou}\ \emph {et~al.}(2018)\citenamefont {Zhou},
  \citenamefont {Song}, \citenamefont {Liu}, \citenamefont {Luan},
  \citenamefont {Wang}, \citenamefont {Sun}, \citenamefont {Jiang},
  \citenamefont {Xiang}, \citenamefont {Chen},\ and\ \citenamefont
  {Du}}]{zhou2018observation}%
  \BibitemOpen
  \bibfield  {author} {\bibinfo {author} {\bibfnamefont {L.}~\bibnamefont
  {Zhou}}, \bibinfo {author} {\bibfnamefont {H.}~\bibnamefont {Song}}, \bibinfo
  {author} {\bibfnamefont {K.}~\bibnamefont {Liu}}, \bibinfo {author}
  {\bibfnamefont {Z.}~\bibnamefont {Luan}}, \bibinfo {author} {\bibfnamefont
  {P.}~\bibnamefont {Wang}}, \bibinfo {author} {\bibfnamefont {L.}~\bibnamefont
  {Sun}}, \bibinfo {author} {\bibfnamefont {S.}~\bibnamefont {Jiang}}, \bibinfo
  {author} {\bibfnamefont {H.}~\bibnamefont {Xiang}}, \bibinfo {author}
  {\bibfnamefont {Y.}~\bibnamefont {Chen}}, \ and\ \bibinfo {author}
  {\bibfnamefont {J.}~\bibnamefont {Du}},\ }\href@noop {} {\bibfield  {journal}
  {\bibinfo  {journal} {Science Advances}\ }\textbf {\bibinfo {volume} {4}},\
  \bibinfo {pages} {eaao3318} (\bibinfo {year} {2018})}\BibitemShut {NoStop}%
\bibitem [{\citenamefont {Maekawa}\ and\ \citenamefont
  {Kimura}(2017)}]{maekawa2017spin}%
  \BibitemOpen
  \bibfield  {author} {\bibinfo {author} {\bibfnamefont {S.}~\bibnamefont
  {Maekawa}}\ and\ \bibinfo {author} {\bibfnamefont {T.}~\bibnamefont
  {Kimura}},\ }\href@noop {} {\emph {\bibinfo {title} {Spin Current}}},\
  Vol.~\bibinfo {volume} {22}\ (\bibinfo  {publisher} {Oxford University
  Press},\ \bibinfo {year} {2017})\BibitemShut {NoStop}%
\bibitem [{\citenamefont {Sinova}\ \emph {et~al.}(2015)\citenamefont {Sinova},
  \citenamefont {Valenzuela}, \citenamefont {Wunderlich}, \citenamefont
  {Back},\ and\ \citenamefont {Jungwirth}}]{sinova2015spin}%
  \BibitemOpen
  \bibfield  {author} {\bibinfo {author} {\bibfnamefont {J.}~\bibnamefont
  {Sinova}}, \bibinfo {author} {\bibfnamefont {S.~O.}\ \bibnamefont
  {Valenzuela}}, \bibinfo {author} {\bibfnamefont {J.}~\bibnamefont
  {Wunderlich}}, \bibinfo {author} {\bibfnamefont {C.}~\bibnamefont {Back}}, \
  and\ \bibinfo {author} {\bibfnamefont {T.}~\bibnamefont {Jungwirth}},\
  }\href@noop {} {\bibfield  {journal} {\bibinfo  {journal} {Reviews of Modern
  Physics}\ }\textbf {\bibinfo {volume} {87}},\ \bibinfo {pages} {1213}
  (\bibinfo {year} {2015})}\BibitemShut {NoStop}%
\bibitem [{\citenamefont {Nakayama}\ \emph {et~al.}(2016)\citenamefont
  {Nakayama}, \citenamefont {Kanno}, \citenamefont {An}, \citenamefont
  {Tashiro}, \citenamefont {Haku}, \citenamefont {Nomura},\ and\ \citenamefont
  {Ando}}]{nakayama2016rashba}%
  \BibitemOpen
  \bibfield  {author} {\bibinfo {author} {\bibfnamefont {H.}~\bibnamefont
  {Nakayama}}, \bibinfo {author} {\bibfnamefont {Y.}~\bibnamefont {Kanno}},
  \bibinfo {author} {\bibfnamefont {H.}~\bibnamefont {An}}, \bibinfo {author}
  {\bibfnamefont {T.}~\bibnamefont {Tashiro}}, \bibinfo {author} {\bibfnamefont
  {S.}~\bibnamefont {Haku}}, \bibinfo {author} {\bibfnamefont {A.}~\bibnamefont
  {Nomura}}, \ and\ \bibinfo {author} {\bibfnamefont {K.}~\bibnamefont
  {Ando}},\ }\href@noop {} {\bibfield  {journal} {\bibinfo  {journal} {Physical
  Review Letters}\ }\textbf {\bibinfo {volume} {117}},\ \bibinfo {pages}
  {116602} (\bibinfo {year} {2016})}\BibitemShut {NoStop}%
\bibitem [{\citenamefont {Isasa}\ \emph {et~al.}(2016)\citenamefont {Isasa},
  \citenamefont {V{\'e}lez}, \citenamefont {Sagasta}, \citenamefont
  {Bedoya-Pinto}, \citenamefont {Dix}, \citenamefont {S{\'a}nchez},
  \citenamefont {Hueso}, \citenamefont {Fontcuberta},\ and\ \citenamefont
  {Casanova}}]{isasa2016spin}%
  \BibitemOpen
  \bibfield  {author} {\bibinfo {author} {\bibfnamefont {M.}~\bibnamefont
  {Isasa}}, \bibinfo {author} {\bibfnamefont {S.}~\bibnamefont {V{\'e}lez}},
  \bibinfo {author} {\bibfnamefont {E.}~\bibnamefont {Sagasta}}, \bibinfo
  {author} {\bibfnamefont {A.}~\bibnamefont {Bedoya-Pinto}}, \bibinfo {author}
  {\bibfnamefont {N.}~\bibnamefont {Dix}}, \bibinfo {author} {\bibfnamefont
  {F.}~\bibnamefont {S{\'a}nchez}}, \bibinfo {author} {\bibfnamefont {L.~E.}\
  \bibnamefont {Hueso}}, \bibinfo {author} {\bibfnamefont {J.}~\bibnamefont
  {Fontcuberta}}, \ and\ \bibinfo {author} {\bibfnamefont {F.}~\bibnamefont
  {Casanova}},\ }\href@noop {} {\bibfield  {journal} {\bibinfo  {journal}
  {Physical Review Applied}\ }\textbf {\bibinfo {volume} {6}},\ \bibinfo
  {pages} {034007} (\bibinfo {year} {2016})}\BibitemShut {NoStop}%
\bibitem [{\citenamefont {Althammer}\ \emph {et~al.}(2013)\citenamefont
  {Althammer}, \citenamefont {Meyer}, \citenamefont {Nakayama}, \citenamefont
  {Schreier}, \citenamefont {Altmannshofer}, \citenamefont {Weiler},
  \citenamefont {Huebl}, \citenamefont {Gepr{\"a}gs}, \citenamefont {Opel},\
  and\ \citenamefont {Gross}}]{althammer2013quantitative}%
  \BibitemOpen
  \bibfield  {author} {\bibinfo {author} {\bibfnamefont {M.}~\bibnamefont
  {Althammer}}, \bibinfo {author} {\bibfnamefont {S.}~\bibnamefont {Meyer}},
  \bibinfo {author} {\bibfnamefont {H.}~\bibnamefont {Nakayama}}, \bibinfo
  {author} {\bibfnamefont {M.}~\bibnamefont {Schreier}}, \bibinfo {author}
  {\bibfnamefont {S.}~\bibnamefont {Altmannshofer}}, \bibinfo {author}
  {\bibfnamefont {M.}~\bibnamefont {Weiler}}, \bibinfo {author} {\bibfnamefont
  {H.}~\bibnamefont {Huebl}}, \bibinfo {author} {\bibfnamefont
  {S.}~\bibnamefont {Gepr{\"a}gs}}, \bibinfo {author} {\bibfnamefont
  {M.}~\bibnamefont {Opel}}, \ and\ \bibinfo {author} {\bibfnamefont
  {R.}~\bibnamefont {Gross}},\ }\href@noop {} {\bibfield  {journal} {\bibinfo
  {journal} {Physical Review B}\ }\textbf {\bibinfo {volume} {87}},\ \bibinfo
  {pages} {224401} (\bibinfo {year} {2013})}\BibitemShut {NoStop}%
\bibitem [{\citenamefont {Huang}\ \emph {et~al.}(2012)\citenamefont {Huang},
  \citenamefont {Fan}, \citenamefont {Qu}, \citenamefont {Chen}, \citenamefont
  {Wang}, \citenamefont {Wu}, \citenamefont {Chen}, \citenamefont {Xiao},\ and\
  \citenamefont {Chien}}]{huang2012transport}%
  \BibitemOpen
  \bibfield  {author} {\bibinfo {author} {\bibfnamefont {S.-Y.}\ \bibnamefont
  {Huang}}, \bibinfo {author} {\bibfnamefont {X.}~\bibnamefont {Fan}}, \bibinfo
  {author} {\bibfnamefont {D.}~\bibnamefont {Qu}}, \bibinfo {author}
  {\bibfnamefont {Y.}~\bibnamefont {Chen}}, \bibinfo {author} {\bibfnamefont
  {W.}~\bibnamefont {Wang}}, \bibinfo {author} {\bibfnamefont {J.}~\bibnamefont
  {Wu}}, \bibinfo {author} {\bibfnamefont {T.}~\bibnamefont {Chen}}, \bibinfo
  {author} {\bibfnamefont {J.}~\bibnamefont {Xiao}}, \ and\ \bibinfo {author}
  {\bibfnamefont {C.}~\bibnamefont {Chien}},\ }\href@noop {} {\bibfield
  {journal} {\bibinfo  {journal} {Physical Review Letters}\ }\textbf {\bibinfo
  {volume} {109}},\ \bibinfo {pages} {107204} (\bibinfo {year}
  {2012})}\BibitemShut {NoStop}%
\bibitem [{\citenamefont {Weiler}\ \emph {et~al.}(2012)\citenamefont {Weiler},
  \citenamefont {Althammer}, \citenamefont {Czeschka}, \citenamefont {Huebl},
  \citenamefont {Wagner}, \citenamefont {Opel}, \citenamefont {Imort},
  \citenamefont {Reiss}, \citenamefont {Thomas}, \citenamefont {Gross},\ and\
  \citenamefont {Goennenwein}}]{weiler2012local}%
  \BibitemOpen
  \bibfield  {author} {\bibinfo {author} {\bibfnamefont {M.}~\bibnamefont
  {Weiler}}, \bibinfo {author} {\bibfnamefont {M.}~\bibnamefont {Althammer}},
  \bibinfo {author} {\bibfnamefont {F.~D.}\ \bibnamefont {Czeschka}}, \bibinfo
  {author} {\bibfnamefont {H.}~\bibnamefont {Huebl}}, \bibinfo {author}
  {\bibfnamefont {M.~S.}\ \bibnamefont {Wagner}}, \bibinfo {author}
  {\bibfnamefont {M.}~\bibnamefont {Opel}}, \bibinfo {author} {\bibfnamefont
  {I.-M.}\ \bibnamefont {Imort}}, \bibinfo {author} {\bibfnamefont
  {G.}~\bibnamefont {Reiss}}, \bibinfo {author} {\bibfnamefont
  {A.}~\bibnamefont {Thomas}}, \bibinfo {author} {\bibfnamefont
  {R.}~\bibnamefont {Gross}}, \ and\ \bibinfo {author} {\bibfnamefont
  {S.~T.~B.}\ \bibnamefont {Goennenwein}},\ }\href {\doibase
  10.1103/PhysRevLett.108.106602} {\bibfield  {journal} {\bibinfo  {journal}
  {Phys. Rev. Lett.}\ }\textbf {\bibinfo {volume} {108}},\ \bibinfo {pages}
  {106602} (\bibinfo {year} {2012})}\BibitemShut {NoStop}%
\bibitem [{\citenamefont {Nakayama}\ \emph {et~al.}(2013)\citenamefont
  {Nakayama}, \citenamefont {Althammer}, \citenamefont {Chen}, \citenamefont
  {Uchida}, \citenamefont {Kajiwara}, \citenamefont {Kikuchi}, \citenamefont
  {Ohtani}, \citenamefont {Gepr\"ags}, \citenamefont {Opel}, \citenamefont
  {Takahashi}, \citenamefont {Gross}, \citenamefont {Bauer}, \citenamefont
  {Goennenwein},\ and\ \citenamefont {Saitoh}}]{nakayama2013spin}%
  \BibitemOpen
  \bibfield  {author} {\bibinfo {author} {\bibfnamefont {H.}~\bibnamefont
  {Nakayama}}, \bibinfo {author} {\bibfnamefont {M.}~\bibnamefont {Althammer}},
  \bibinfo {author} {\bibfnamefont {Y.-T.}\ \bibnamefont {Chen}}, \bibinfo
  {author} {\bibfnamefont {K.}~\bibnamefont {Uchida}}, \bibinfo {author}
  {\bibfnamefont {Y.}~\bibnamefont {Kajiwara}}, \bibinfo {author}
  {\bibfnamefont {D.}~\bibnamefont {Kikuchi}}, \bibinfo {author} {\bibfnamefont
  {T.}~\bibnamefont {Ohtani}}, \bibinfo {author} {\bibfnamefont
  {S.}~\bibnamefont {Gepr\"ags}}, \bibinfo {author} {\bibfnamefont
  {M.}~\bibnamefont {Opel}}, \bibinfo {author} {\bibfnamefont {S.}~\bibnamefont
  {Takahashi}}, \bibinfo {author} {\bibfnamefont {R.}~\bibnamefont {Gross}},
  \bibinfo {author} {\bibfnamefont {G.~E.~W.}\ \bibnamefont {Bauer}}, \bibinfo
  {author} {\bibfnamefont {S.~T.~B.}\ \bibnamefont {Goennenwein}}, \ and\
  \bibinfo {author} {\bibfnamefont {E.}~\bibnamefont {Saitoh}},\ }\href
  {\doibase 10.1103/PhysRevLett.110.206601} {\bibfield  {journal} {\bibinfo
  {journal} {Phys. Rev. Lett.}\ }\textbf {\bibinfo {volume} {110}},\ \bibinfo
  {pages} {206601} (\bibinfo {year} {2013})}\BibitemShut {NoStop}%
\bibitem [{\citenamefont {Avci}\ \emph {et~al.}(2015)\citenamefont {Avci},
  \citenamefont {Garello}, \citenamefont {Ghosh}, \citenamefont {Gabureac},
  \citenamefont {Alvarado},\ and\ \citenamefont
  {Gambardella}}]{avci2015unidirectional}%
  \BibitemOpen
  \bibfield  {author} {\bibinfo {author} {\bibfnamefont {C.~O.}\ \bibnamefont
  {Avci}}, \bibinfo {author} {\bibfnamefont {K.}~\bibnamefont {Garello}},
  \bibinfo {author} {\bibfnamefont {A.}~\bibnamefont {Ghosh}}, \bibinfo
  {author} {\bibfnamefont {M.}~\bibnamefont {Gabureac}}, \bibinfo {author}
  {\bibfnamefont {S.~F.}\ \bibnamefont {Alvarado}}, \ and\ \bibinfo {author}
  {\bibfnamefont {P.}~\bibnamefont {Gambardella}},\ }\href@noop {} {\bibfield
  {journal} {\bibinfo  {journal} {Nature Physics}\ }\textbf {\bibinfo {volume}
  {11}},\ \bibinfo {pages} {570} (\bibinfo {year} {2015})}\BibitemShut
  {NoStop}%
\bibitem [{\citenamefont {Hahn}\ \emph {et~al.}(2013)\citenamefont {Hahn},
  \citenamefont {De~Loubens}, \citenamefont {Klein}, \citenamefont {Viret},
  \citenamefont {Naletov},\ and\ \citenamefont
  {Youssef}}]{hahn2013comparative}%
  \BibitemOpen
  \bibfield  {author} {\bibinfo {author} {\bibfnamefont {C.}~\bibnamefont
  {Hahn}}, \bibinfo {author} {\bibfnamefont {G.}~\bibnamefont {De~Loubens}},
  \bibinfo {author} {\bibfnamefont {O.}~\bibnamefont {Klein}}, \bibinfo
  {author} {\bibfnamefont {M.}~\bibnamefont {Viret}}, \bibinfo {author}
  {\bibfnamefont {V.~V.}\ \bibnamefont {Naletov}}, \ and\ \bibinfo {author}
  {\bibfnamefont {J.~B.}\ \bibnamefont {Youssef}},\ }\href@noop {} {\bibfield
  {journal} {\bibinfo  {journal} {Physical Review B}\ }\textbf {\bibinfo
  {volume} {87}},\ \bibinfo {pages} {174417} (\bibinfo {year}
  {2013})}\BibitemShut {NoStop}%
\bibitem [{\citenamefont {Dejene}\ \emph {et~al.}(2015)\citenamefont {Dejene},
  \citenamefont {Vlietstra}, \citenamefont {Luc}, \citenamefont {Waintal},
  \citenamefont {Youssef},\ and\ \citenamefont {Van~Wees}}]{dejene2015control}%
  \BibitemOpen
  \bibfield  {author} {\bibinfo {author} {\bibfnamefont {F.}~\bibnamefont
  {Dejene}}, \bibinfo {author} {\bibfnamefont {N.}~\bibnamefont {Vlietstra}},
  \bibinfo {author} {\bibfnamefont {D.}~\bibnamefont {Luc}}, \bibinfo {author}
  {\bibfnamefont {X.}~\bibnamefont {Waintal}}, \bibinfo {author} {\bibfnamefont
  {J.~B.}\ \bibnamefont {Youssef}}, \ and\ \bibinfo {author} {\bibfnamefont
  {B.}~\bibnamefont {Van~Wees}},\ }\href@noop {} {\bibfield  {journal}
  {\bibinfo  {journal} {Physical Review B}\ }\textbf {\bibinfo {volume} {91}},\
  \bibinfo {pages} {100404} (\bibinfo {year} {2015})}\BibitemShut {NoStop}%
\bibitem [{\citenamefont {Chen}\ \emph {et~al.}(2013)\citenamefont {Chen},
  \citenamefont {Takahashi}, \citenamefont {Nakayama}, \citenamefont
  {Althammer}, \citenamefont {Goennenwein}, \citenamefont {Saitoh},\ and\
  \citenamefont {Bauer}}]{chen2013theory}%
  \BibitemOpen
  \bibfield  {author} {\bibinfo {author} {\bibfnamefont {Y.-T.}\ \bibnamefont
  {Chen}}, \bibinfo {author} {\bibfnamefont {S.}~\bibnamefont {Takahashi}},
  \bibinfo {author} {\bibfnamefont {H.}~\bibnamefont {Nakayama}}, \bibinfo
  {author} {\bibfnamefont {M.}~\bibnamefont {Althammer}}, \bibinfo {author}
  {\bibfnamefont {S.~T.}\ \bibnamefont {Goennenwein}}, \bibinfo {author}
  {\bibfnamefont {E.}~\bibnamefont {Saitoh}}, \ and\ \bibinfo {author}
  {\bibfnamefont {G.~E.}\ \bibnamefont {Bauer}},\ }\href@noop {} {\bibfield
  {journal} {\bibinfo  {journal} {Physical Review B}\ }\textbf {\bibinfo
  {volume} {87}},\ \bibinfo {pages} {144411} (\bibinfo {year}
  {2013})}\BibitemShut {NoStop}%
\bibitem [{\citenamefont {Jia}\ \emph {et~al.}(2011)\citenamefont {Jia},
  \citenamefont {Liu}, \citenamefont {Xia},\ and\ \citenamefont
  {Bauer}}]{jia2011spin}%
  \BibitemOpen
  \bibfield  {author} {\bibinfo {author} {\bibfnamefont {X.}~\bibnamefont
  {Jia}}, \bibinfo {author} {\bibfnamefont {K.}~\bibnamefont {Liu}}, \bibinfo
  {author} {\bibfnamefont {K.}~\bibnamefont {Xia}}, \ and\ \bibinfo {author}
  {\bibfnamefont {G.~E.}\ \bibnamefont {Bauer}},\ }\href@noop {} {\bibfield
  {journal} {\bibinfo  {journal} {EPL (Europhysics Letters)}\ }\textbf
  {\bibinfo {volume} {96}},\ \bibinfo {pages} {17005} (\bibinfo {year}
  {2011})}\BibitemShut {NoStop}%
\bibitem [{\citenamefont {Carva}\ and\ \citenamefont
  {Turek}(2007)}]{carva2007spin}%
  \BibitemOpen
  \bibfield  {author} {\bibinfo {author} {\bibfnamefont {K.}~\bibnamefont
  {Carva}}\ and\ \bibinfo {author} {\bibfnamefont {I.}~\bibnamefont {Turek}},\
  }\href@noop {} {\bibfield  {journal} {\bibinfo  {journal} {Physical Review
  B}\ }\textbf {\bibinfo {volume} {76}},\ \bibinfo {pages} {104409} (\bibinfo
  {year} {2007})}\BibitemShut {NoStop}%
\bibitem [{\citenamefont {Zhang}\ \emph {et~al.}(2011)\citenamefont {Zhang},
  \citenamefont {Hikino},\ and\ \citenamefont {Yunoki}}]{zhang2011first}%
  \BibitemOpen
  \bibfield  {author} {\bibinfo {author} {\bibfnamefont {Q.}~\bibnamefont
  {Zhang}}, \bibinfo {author} {\bibfnamefont {S.-i.}\ \bibnamefont {Hikino}}, \
  and\ \bibinfo {author} {\bibfnamefont {S.}~\bibnamefont {Yunoki}},\
  }\href@noop {} {\bibfield  {journal} {\bibinfo  {journal} {Applied Physics
  Letters}\ }\textbf {\bibinfo {volume} {99}},\ \bibinfo {pages} {172105}
  (\bibinfo {year} {2011})}\BibitemShut {NoStop}%
\bibitem [{\citenamefont {Xia}\ \emph {et~al.}(2002)\citenamefont {Xia},
  \citenamefont {Kelly}, \citenamefont {Bauer}, \citenamefont {Brataas},\ and\
  \citenamefont {Turek}}]{xia2002spin}%
  \BibitemOpen
  \bibfield  {author} {\bibinfo {author} {\bibfnamefont {K.}~\bibnamefont
  {Xia}}, \bibinfo {author} {\bibfnamefont {P.~J.}\ \bibnamefont {Kelly}},
  \bibinfo {author} {\bibfnamefont {G.}~\bibnamefont {Bauer}}, \bibinfo
  {author} {\bibfnamefont {A.}~\bibnamefont {Brataas}}, \ and\ \bibinfo
  {author} {\bibfnamefont {I.}~\bibnamefont {Turek}},\ }\href@noop {}
  {\bibfield  {journal} {\bibinfo  {journal} {Physical Review B}\ }\textbf
  {\bibinfo {volume} {65}},\ \bibinfo {pages} {220401} (\bibinfo {year}
  {2002})}\BibitemShut {NoStop}%
\bibitem [{\citenamefont {Dolui}\ \emph {et~al.}(2019)\citenamefont {Dolui},
  \citenamefont {Bajpai},\ and\ \citenamefont {Nikolic}}]{dolui2019spin}%
  \BibitemOpen
  \bibfield  {author} {\bibinfo {author} {\bibfnamefont {K.}~\bibnamefont
  {Dolui}}, \bibinfo {author} {\bibfnamefont {U.}~\bibnamefont {Bajpai}}, \
  and\ \bibinfo {author} {\bibfnamefont {B.~K.}\ \bibnamefont {Nikolic}},\
  }\href@noop {} {\ ,\ \bibinfo {pages} {arXiv:1905.01299} (\bibinfo {year}
  {2019})}\BibitemShut {NoStop}%
\bibitem [{\citenamefont {Meyer}\ \emph {et~al.}(2014)\citenamefont {Meyer},
  \citenamefont {Althammer}, \citenamefont {Gepr{\"a}gs}, \citenamefont {Opel},
  \citenamefont {Gross},\ and\ \citenamefont
  {Goennenwein}}]{meyer2014temperature}%
  \BibitemOpen
  \bibfield  {author} {\bibinfo {author} {\bibfnamefont {S.}~\bibnamefont
  {Meyer}}, \bibinfo {author} {\bibfnamefont {M.}~\bibnamefont {Althammer}},
  \bibinfo {author} {\bibfnamefont {S.}~\bibnamefont {Gepr{\"a}gs}}, \bibinfo
  {author} {\bibfnamefont {M.}~\bibnamefont {Opel}}, \bibinfo {author}
  {\bibfnamefont {R.}~\bibnamefont {Gross}}, \ and\ \bibinfo {author}
  {\bibfnamefont {S.~T.}\ \bibnamefont {Goennenwein}},\ }\href@noop {}
  {\bibfield  {journal} {\bibinfo  {journal} {Applied Physics Letters}\
  }\textbf {\bibinfo {volume} {104}},\ \bibinfo {pages} {242411} (\bibinfo
  {year} {2014})}\BibitemShut {NoStop}%
\bibitem [{\citenamefont {V{\'e}lez}\ \emph {et~al.}(2018)\citenamefont
  {V{\'e}lez}, \citenamefont {Golovach}, \citenamefont {Gomez-Perez},
  \citenamefont {Bui}, \citenamefont {Rivadulla}, \citenamefont {Hueso},
  \citenamefont {Bergeret},\ and\ \citenamefont {Casanova}}]{velez2018spin}%
  \BibitemOpen
  \bibfield  {author} {\bibinfo {author} {\bibfnamefont {S.}~\bibnamefont
  {V{\'e}lez}}, \bibinfo {author} {\bibfnamefont {V.~N.}\ \bibnamefont
  {Golovach}}, \bibinfo {author} {\bibfnamefont {J.~M.}\ \bibnamefont
  {Gomez-Perez}}, \bibinfo {author} {\bibfnamefont {C.~T.}\ \bibnamefont
  {Bui}}, \bibinfo {author} {\bibfnamefont {F.}~\bibnamefont {Rivadulla}},
  \bibinfo {author} {\bibfnamefont {L.~E.}\ \bibnamefont {Hueso}}, \bibinfo
  {author} {\bibfnamefont {F.~S.}\ \bibnamefont {Bergeret}}, \ and\ \bibinfo
  {author} {\bibfnamefont {F.}~\bibnamefont {Casanova}},\ }\href@noop {} {\ ,\
  \bibinfo {pages} {arXiv:1805.11225} (\bibinfo {year} {2018})}\BibitemShut
  {NoStop}%
\bibitem [{\citenamefont {Das}\ \emph {et~al.}(2019)\citenamefont {Das},
  \citenamefont {Dejene}, \citenamefont {van Wees},\ and\ \citenamefont
  {Vera-Marun}}]{das2019temperature}%
  \BibitemOpen
  \bibfield  {author} {\bibinfo {author} {\bibfnamefont {K.}~\bibnamefont
  {Das}}, \bibinfo {author} {\bibfnamefont {F.}~\bibnamefont {Dejene}},
  \bibinfo {author} {\bibfnamefont {B.}~\bibnamefont {van Wees}}, \ and\
  \bibinfo {author} {\bibfnamefont {I.}~\bibnamefont {Vera-Marun}},\
  }\href@noop {} {\bibfield  {journal} {\bibinfo  {journal} {Applied Physics
  Letters}\ }\textbf {\bibinfo {volume} {114}},\ \bibinfo {pages} {072405}
  (\bibinfo {year} {2019})}\BibitemShut {NoStop}%
\bibitem [{\citenamefont {Dyakonov}(2007)}]{dyakonov2007magnetoresistance}%
  \BibitemOpen
  \bibfield  {author} {\bibinfo {author} {\bibfnamefont {M.}~\bibnamefont
  {Dyakonov}},\ }\href@noop {} {\bibfield  {journal} {\bibinfo  {journal}
  {Physical Review Letters}\ }\textbf {\bibinfo {volume} {99}},\ \bibinfo
  {pages} {126601} (\bibinfo {year} {2007})}\BibitemShut {NoStop}%
\bibitem [{\citenamefont {V{\'e}lez}\ \emph {et~al.}(2016)\citenamefont
  {V{\'e}lez}, \citenamefont {Golovach}, \citenamefont {Bedoya-Pinto},
  \citenamefont {Isasa}, \citenamefont {Sagasta}, \citenamefont {Abadia},
  \citenamefont {Rogero}, \citenamefont {Hueso}, \citenamefont {Bergeret},\
  and\ \citenamefont {Casanova}}]{velez2016hanle}%
  \BibitemOpen
  \bibfield  {author} {\bibinfo {author} {\bibfnamefont {S.}~\bibnamefont
  {V{\'e}lez}}, \bibinfo {author} {\bibfnamefont {V.~N.}\ \bibnamefont
  {Golovach}}, \bibinfo {author} {\bibfnamefont {A.}~\bibnamefont
  {Bedoya-Pinto}}, \bibinfo {author} {\bibfnamefont {M.}~\bibnamefont {Isasa}},
  \bibinfo {author} {\bibfnamefont {E.}~\bibnamefont {Sagasta}}, \bibinfo
  {author} {\bibfnamefont {M.}~\bibnamefont {Abadia}}, \bibinfo {author}
  {\bibfnamefont {C.}~\bibnamefont {Rogero}}, \bibinfo {author} {\bibfnamefont
  {L.~E.}\ \bibnamefont {Hueso}}, \bibinfo {author} {\bibfnamefont {F.~S.}\
  \bibnamefont {Bergeret}}, \ and\ \bibinfo {author} {\bibfnamefont
  {F.}~\bibnamefont {Casanova}},\ }\href@noop {} {\bibfield  {journal}
  {\bibinfo  {journal} {Physical Review Letters}\ }\textbf {\bibinfo {volume}
  {116}},\ \bibinfo {pages} {016603} (\bibinfo {year} {2016})}\BibitemShut
  {NoStop}%
\bibitem [{Note1()}]{Note1}%
  \BibitemOpen
  \bibinfo {note} {It is important to remark that the coupling in Eq.~(\ref
  {eq:Vsd}) acts more efficiently when the spin $\protect \bm {S}_i$ is
  embedded in the metal as compared to the case when it is at the surface and
  interacts only with the tail of the electron wave function appearing in
  $\protect \bm {s}(\protect \bm {r}_i)$. We should, therefore, reduce
  $J_{\protect \textit {sd}}$ in Eq.~(\ref {eq:Vsd}) by a factor $n(z>\lambda
  _F)/n(z=0)$, where $n(\protect \bm {r})$ is the average charge density in the
  metal and $\lambda _{F}$ is the Fermi wave length. However, this suppression
  factor is expected to be on the order of unity in well-coupled systems, for
  which the local moments at the surface form bonds with the metal. We absorb
  this suppression factor into $J_{sd}$ hereafter.}\BibitemShut {Stop}%
\bibitem [{\citenamefont {Hao}\ \emph {et~al.}(1990)\citenamefont {Hao},
  \citenamefont {Moodera},\ and\ \citenamefont {Meservey}}]{hao1990spin}%
  \BibitemOpen
  \bibfield  {author} {\bibinfo {author} {\bibfnamefont {X.}~\bibnamefont
  {Hao}}, \bibinfo {author} {\bibfnamefont {J.}~\bibnamefont {Moodera}}, \ and\
  \bibinfo {author} {\bibfnamefont {R.}~\bibnamefont {Meservey}},\ }\href@noop
  {} {\bibfield  {journal} {\bibinfo  {journal} {Physical Review B}\ }\textbf
  {\bibinfo {volume} {42}},\ \bibinfo {pages} {8235} (\bibinfo {year}
  {1990})}\BibitemShut {NoStop}%
\bibitem [{\citenamefont {Strambini}\ \emph {et~al.}(2017)\citenamefont
  {Strambini}, \citenamefont {Golovach}, \citenamefont {De~Simoni},
  \citenamefont {Moodera}, \citenamefont {Bergeret},\ and\ \citenamefont
  {Giazotto}}]{strambini2017revealing}%
  \BibitemOpen
  \bibfield  {author} {\bibinfo {author} {\bibfnamefont {E.}~\bibnamefont
  {Strambini}}, \bibinfo {author} {\bibfnamefont {V.}~\bibnamefont {Golovach}},
  \bibinfo {author} {\bibfnamefont {G.}~\bibnamefont {De~Simoni}}, \bibinfo
  {author} {\bibfnamefont {J.}~\bibnamefont {Moodera}}, \bibinfo {author}
  {\bibfnamefont {F.}~\bibnamefont {Bergeret}}, \ and\ \bibinfo {author}
  {\bibfnamefont {F.}~\bibnamefont {Giazotto}},\ }\href@noop {} {\bibfield
  {journal} {\bibinfo  {journal} {Physical Review Materials}\ }\textbf
  {\bibinfo {volume} {1}},\ \bibinfo {pages} {054402} (\bibinfo {year}
  {2017})}\BibitemShut {NoStop}%
\bibitem [{\citenamefont {Slichter}(2013)}]{slichter2013principles}%
  \BibitemOpen
  \bibfield  {author} {\bibinfo {author} {\bibfnamefont {C.~P.}\ \bibnamefont
  {Slichter}},\ }\href@noop {} {\emph {\bibinfo {title} {Principles of magnetic
  resonance}}},\ Vol.~\bibinfo {volume} {1}\ (\bibinfo  {publisher} {Springer
  Science \& Business Media},\ \bibinfo {year} {2013})\BibitemShut {NoStop}%
\bibitem [{\citenamefont {Brataas}\ \emph {et~al.}(2001)\citenamefont
  {Brataas}, \citenamefont {Nazarov},\ and\ \citenamefont
  {Bauer}}]{brataas2001spin}%
  \BibitemOpen
  \bibfield  {author} {\bibinfo {author} {\bibfnamefont {A.}~\bibnamefont
  {Brataas}}, \bibinfo {author} {\bibfnamefont {Y.~V.}\ \bibnamefont
  {Nazarov}}, \ and\ \bibinfo {author} {\bibfnamefont {G.~E.}\ \bibnamefont
  {Bauer}},\ }\href@noop {} {\bibfield  {journal} {\bibinfo  {journal} {The
  European Physical Journal B-Condensed Matter and Complex Systems}\ }\textbf
  {\bibinfo {volume} {22}},\ \bibinfo {pages} {99} (\bibinfo {year}
  {2001})}\BibitemShut {NoStop}%
\bibitem [{\citenamefont {Wahl}\ \emph {et~al.}(2007)\citenamefont {Wahl},
  \citenamefont {Simon}, \citenamefont {Diekh{\"o}ner}, \citenamefont
  {Stepanyuk}, \citenamefont {Bruno}, \citenamefont {Schneider},\ and\
  \citenamefont {Kern}}]{wahl2007exchange}%
  \BibitemOpen
  \bibfield  {author} {\bibinfo {author} {\bibfnamefont {P.}~\bibnamefont
  {Wahl}}, \bibinfo {author} {\bibfnamefont {P.}~\bibnamefont {Simon}},
  \bibinfo {author} {\bibfnamefont {L.}~\bibnamefont {Diekh{\"o}ner}}, \bibinfo
  {author} {\bibfnamefont {V.}~\bibnamefont {Stepanyuk}}, \bibinfo {author}
  {\bibfnamefont {P.}~\bibnamefont {Bruno}}, \bibinfo {author} {\bibfnamefont
  {M.}~\bibnamefont {Schneider}}, \ and\ \bibinfo {author} {\bibfnamefont
  {K.}~\bibnamefont {Kern}},\ }\href@noop {} {\bibfield  {journal} {\bibinfo
  {journal} {Physical Review Letters}\ }\textbf {\bibinfo {volume} {98}},\
  \bibinfo {pages} {056601} (\bibinfo {year} {2007})}\BibitemShut {NoStop}%
\bibitem [{\citenamefont {Vlietstra}\ \emph {et~al.}(2013)\citenamefont
  {Vlietstra}, \citenamefont {Shan}, \citenamefont {Castel}, \citenamefont
  {Ben~Youssef}, \citenamefont {Bauer},\ and\ \citenamefont
  {Van~Wees}}]{vlietstra2013exchange}%
  \BibitemOpen
  \bibfield  {author} {\bibinfo {author} {\bibfnamefont {N.}~\bibnamefont
  {Vlietstra}}, \bibinfo {author} {\bibfnamefont {J.}~\bibnamefont {Shan}},
  \bibinfo {author} {\bibfnamefont {V.}~\bibnamefont {Castel}}, \bibinfo
  {author} {\bibfnamefont {J.}~\bibnamefont {Ben~Youssef}}, \bibinfo {author}
  {\bibfnamefont {G.}~\bibnamefont {Bauer}}, \ and\ \bibinfo {author}
  {\bibfnamefont {B.}~\bibnamefont {Van~Wees}},\ }\href@noop {} {\bibfield
  {journal} {\bibinfo  {journal} {Applied Physics Letters}\ }\textbf {\bibinfo
  {volume} {103}},\ \bibinfo {pages} {032401} (\bibinfo {year}
  {2013})}\BibitemShut {NoStop}%
\bibitem [{\citenamefont {Isasa}\ \emph {et~al.}(2014)\citenamefont {Isasa},
  \citenamefont {Bedoya-Pinto}, \citenamefont {V{\'e}lez}, \citenamefont
  {Golmar}, \citenamefont {S{\'a}nchez}, \citenamefont {Hueso}, \citenamefont
  {Fontcuberta},\ and\ \citenamefont {Casanova}}]{isasa2014spin}%
  \BibitemOpen
  \bibfield  {author} {\bibinfo {author} {\bibfnamefont {M.}~\bibnamefont
  {Isasa}}, \bibinfo {author} {\bibfnamefont {A.}~\bibnamefont {Bedoya-Pinto}},
  \bibinfo {author} {\bibfnamefont {S.}~\bibnamefont {V{\'e}lez}}, \bibinfo
  {author} {\bibfnamefont {F.}~\bibnamefont {Golmar}}, \bibinfo {author}
  {\bibfnamefont {F.}~\bibnamefont {S{\'a}nchez}}, \bibinfo {author}
  {\bibfnamefont {L.~E.}\ \bibnamefont {Hueso}}, \bibinfo {author}
  {\bibfnamefont {J.}~\bibnamefont {Fontcuberta}}, \ and\ \bibinfo {author}
  {\bibfnamefont {F.}~\bibnamefont {Casanova}},\ }\href@noop {} {\bibfield
  {journal} {\bibinfo  {journal} {Applied Physics Letters}\ }\textbf {\bibinfo
  {volume} {105}},\ \bibinfo {pages} {142402} (\bibinfo {year}
  {2014})}\BibitemShut {NoStop}%
\bibitem [{\citenamefont {Lammel}\ \emph {et~al.}(2019)\citenamefont {Lammel},
  \citenamefont {Schlitz}, \citenamefont {Geishendorf}, \citenamefont
  {Makarov}, \citenamefont {Kosub}, \citenamefont {Fabretti}, \citenamefont
  {Reichlova}, \citenamefont {Huebner}, \citenamefont {Nielsch}, \citenamefont
  {Thomas},\ and\ \citenamefont {Goennenwein}}]{lammel2019spin}%
  \BibitemOpen
  \bibfield  {author} {\bibinfo {author} {\bibfnamefont {M.}~\bibnamefont
  {Lammel}}, \bibinfo {author} {\bibfnamefont {R.}~\bibnamefont {Schlitz}},
  \bibinfo {author} {\bibfnamefont {K.}~\bibnamefont {Geishendorf}}, \bibinfo
  {author} {\bibfnamefont {D.}~\bibnamefont {Makarov}}, \bibinfo {author}
  {\bibfnamefont {T.}~\bibnamefont {Kosub}}, \bibinfo {author} {\bibfnamefont
  {S.}~\bibnamefont {Fabretti}}, \bibinfo {author} {\bibfnamefont
  {H.}~\bibnamefont {Reichlova}}, \bibinfo {author} {\bibfnamefont
  {R.}~\bibnamefont {Huebner}}, \bibinfo {author} {\bibfnamefont
  {K.}~\bibnamefont {Nielsch}}, \bibinfo {author} {\bibfnamefont
  {A.}~\bibnamefont {Thomas}}, \ and\ \bibinfo {author} {\bibfnamefont {S.~T.}\
  \bibnamefont {Goennenwein}},\ }\href@noop {} {\ ,\ \bibinfo {pages}
  {arXiv:1901.09986} (\bibinfo {year} {2019})}\BibitemShut {NoStop}%
\bibitem [{\citenamefont {Koichi}\ and\ \citenamefont
  {et~al.}()}]{koichi2019paramagneticSMR}%
  \BibitemOpen
  \bibfield  {author} {\bibinfo {author} {\bibfnamefont {O.}~\bibnamefont
  {Koichi}}\ and\ \bibinfo {author} {\bibnamefont {et~al.}},\ }\href@noop {}
  {\bibinfo  {journal} {In preparation.}\ }\BibitemShut {NoStop}%
\bibitem [{\citenamefont {Fontcuberta}\ \emph {et~al.}(2019)\citenamefont
  {Fontcuberta}, \citenamefont {Vasili}, \citenamefont {G{\`a}zquez},\ and\
  \citenamefont {Casanova}}]{fontcuberta2019role}%
  \BibitemOpen
\bibfield  {journal} {  }\bibfield  {author} {\bibinfo {author} {\bibfnamefont
  {J.}~\bibnamefont {Fontcuberta}}, \bibinfo {author} {\bibfnamefont {H.~B.}\
  \bibnamefont {Vasili}}, \bibinfo {author} {\bibfnamefont {J.}~\bibnamefont
  {G{\`a}zquez}}, \ and\ \bibinfo {author} {\bibfnamefont {F.}~\bibnamefont
  {Casanova}},\ }\href@noop {} {\bibfield  {journal} {\bibinfo  {journal}
  {Advanced Materials Interfaces}\ ,\ \bibinfo {pages} {1900475}} (\bibinfo
  {year} {2019})}\BibitemShut {NoStop}%
\bibitem [{\citenamefont {Myers}\ \emph {et~al.}(2005)\citenamefont {Myers},
  \citenamefont {Poggio}, \citenamefont {Stern}, \citenamefont {Gossard},\ and\
  \citenamefont {Awschalom}}]{myers2005antiferromagnetic}%
  \BibitemOpen
  \bibfield  {author} {\bibinfo {author} {\bibfnamefont {R.}~\bibnamefont
  {Myers}}, \bibinfo {author} {\bibfnamefont {M.}~\bibnamefont {Poggio}},
  \bibinfo {author} {\bibfnamefont {N.}~\bibnamefont {Stern}}, \bibinfo
  {author} {\bibfnamefont {A.}~\bibnamefont {Gossard}}, \ and\ \bibinfo
  {author} {\bibfnamefont {D.}~\bibnamefont {Awschalom}},\ }\href@noop {}
  {\bibfield  {journal} {\bibinfo  {journal} {Physical Review Letters}\
  }\textbf {\bibinfo {volume} {95}},\ \bibinfo {pages} {017204} (\bibinfo
  {year} {2005})}\BibitemShut {NoStop}%
\bibitem [{\citenamefont {M{\"u}ller}\ \emph {et~al.}(2009)\citenamefont
  {M{\"u}ller}, \citenamefont {Miao},\ and\ \citenamefont
  {Moodera}}]{muller2009thickness}%
  \BibitemOpen
  \bibfield  {author} {\bibinfo {author} {\bibfnamefont {M.}~\bibnamefont
  {M{\"u}ller}}, \bibinfo {author} {\bibfnamefont {G.-X.}\ \bibnamefont
  {Miao}}, \ and\ \bibinfo {author} {\bibfnamefont {J.~S.}\ \bibnamefont
  {Moodera}},\ }\href@noop {} {\bibfield  {journal} {\bibinfo  {journal}
  {Journal of Applied Physics}\ }\textbf {\bibinfo {volume} {105}},\ \bibinfo
  {pages} {07C917} (\bibinfo {year} {2009})}\BibitemShut {NoStop}%
\bibitem [{\citenamefont {Wei}\ \emph {et~al.}(2016)\citenamefont {Wei},
  \citenamefont {Lee}, \citenamefont {Lemaitre}, \citenamefont {Pinel},
  \citenamefont {Cutaia}, \citenamefont {Cha}, \citenamefont {Katmis},
  \citenamefont {Zhu}, \citenamefont {Heiman}, \citenamefont {Hone},
  \citenamefont {Moodera},\ and\ \citenamefont {Chen}}]{wei2016strong}%
  \BibitemOpen
  \bibfield  {author} {\bibinfo {author} {\bibfnamefont {P.}~\bibnamefont
  {Wei}}, \bibinfo {author} {\bibfnamefont {S.}~\bibnamefont {Lee}}, \bibinfo
  {author} {\bibfnamefont {F.}~\bibnamefont {Lemaitre}}, \bibinfo {author}
  {\bibfnamefont {L.}~\bibnamefont {Pinel}}, \bibinfo {author} {\bibfnamefont
  {D.}~\bibnamefont {Cutaia}}, \bibinfo {author} {\bibfnamefont
  {W.}~\bibnamefont {Cha}}, \bibinfo {author} {\bibfnamefont {F.}~\bibnamefont
  {Katmis}}, \bibinfo {author} {\bibfnamefont {Y.}~\bibnamefont {Zhu}},
  \bibinfo {author} {\bibfnamefont {D.}~\bibnamefont {Heiman}}, \bibinfo
  {author} {\bibfnamefont {J.}~\bibnamefont {Hone}}, \bibinfo {author}
  {\bibfnamefont {J.~S.}\ \bibnamefont {Moodera}}, \ and\ \bibinfo {author}
  {\bibfnamefont {C.-T.}\ \bibnamefont {Chen}},\ }\href@noop {} {\bibfield
  {journal} {\bibinfo  {journal} {Nature Materials}\ }\textbf {\bibinfo
  {volume} {15}},\ \bibinfo {pages} {711} (\bibinfo {year} {2016})}\BibitemShut
  {NoStop}%
\end{thebibliography}

%

\end{document}